\documentclass[AMA,STIX1COL]{WileyNJD-v2}
\articletype{Article Type}%
\received{26 April 2016} \revised{6 June 2016} \accepted{6 June 2016}

\newcommand{\la}{\lambda}

\newcommand{\al }{\lambda}
\newcommand{\Al}{\Lambda}

\newcommand{\axis}{axis~}

\begin{document}
\title{Analysis of time-to-event for observational studies: Guidance to the use of intensity models}

\author[1]{Per Kragh Andersen}

\author[2]{Maja Pohar Perme*}

\author[3]{Hans C. van Houwelingen}

\author[4]{Richard J. Cook}

\author[5]{Pierre Joly}
\author[1]{Torben Martinussen}
\author[6]{Jeremy M.G. Taylor}
\author[7]{Michal Abrahamowicz}
\author[8]{Terry M. Therneau}

\author{for the
 STRATOS TG8 topic group}

\address[1]{\orgdiv{Section of Biostatistics}, \orgname{University of Copenhagen}, \country{Denmark}}

\address[2]{\orgdiv{Department of Biostatistics and Medical Informatics}, \orgname{Medical faculty, University of Ljubljana},  \country{Slovenia}}

\address[3]{\orgdiv{Department of Biomedical Data Sciences}, \orgname{Leiden University}, \country{The Netherlands}}

\address[4]{\orgdiv{Department of Statistics and Actuarial Science}, \orgname{University of Waterloo}, \country{Canada}}

\address[5]{\orgdiv{Inserm, ISPED, Bordeaux Populations Health Research Center}, \orgname{University of Bordeaux}, \country{France}}

\address[6]{\orgdiv{Department of Biostatistics}, \orgname{University of Michigan}, \country{USA}}

\address[7]{\orgdiv{Department of Epidemiology, Biostatistics and Occupational Health}, \orgname{McGill University}, \orgaddress{\state{Montreal}, \country{Canada}}}

\address[8]{\orgdiv{Division of Biomedical Statistics and Informatics}, \orgname{ Mayo Clinic, Rochester}, \country{USA}}

\corres{*Maja Pohar Perme,  \email{maja.pohar@mf.uni-lj.si}}

\abstract[Summary]{This paper provides guidance for researchers with some
mathematical background on the conduct of time-to-event analysis in
observational studies based on intensity (hazard) models. Discussions
of basic concepts like time axis, event definition and censoring are
given. Hazard models are
introduced, with special emphasis on the Cox proportional hazards
regression model. We provide check lists that may be useful both when
fitting the model and assessing its goodness of fit and when
interpreting the results. Special attention is paid to how to avoid
problems with immortal time bias by introducing time-dependent
covariates. We discuss prediction based on hazard models and
difficulties when attempting to draw proper causal conclusions from
such models. Finally, we present a series of examples where the methods
and check lists are exemplified. Computational details and
implementation using the freely available {\tt R} software are
documented in Supplementary Material. The paper was prepared as part of the STRATOS
initiative.}



\maketitle

\section{Introduction}
\label{sec:intro}

Methods for survival, or time-to-event, analysis are frequently used in epidemiological and clinical
studies of human health. The more
than 30,000 Pubmed citations for the Cox proportional hazards model
alone attest to the critical role of such methods in modern health research. Most of the observable health outcomes, such as disease
onset, progression, cure or death, are the result of
the evolution of relevant biological systems resulting from a natural aging process or the
effects of exposures and treatments that may accumulate over time; hence a
time-to-event paradigm provides a natural framework for their
analyses. Accordingly, biostatisticians working in medical
research are very likely to encounter problems requiring time-to-event analyses, even if their training and
interests lie in different areas of statistical research. Time-to-event data typically feature
 particular challenges related to, among other things, censored
observations and
changes over time in the absolute and/or relative risks, as well as in
the values of the predictors. To further complicate matters, there
are several issues in survival analysis for which no clear consensus,
or published guidelines exist.
The lack of clear guidance on how to address these challenges
 may explain why many
published applications involving survival analysis have important
weaknesses \citep[e.g.][]{Altman95}.

These considerations motivated us to create the `Survival Analysis'
topic group within the STRATOS (STRengthening Analytical Thinking for
Observational Studies) initiative. The over-arching aim of the STRATOS initiative is to provide guidance for accurate and efficient
analyses in different areas of statistics relevant for observational
(non-randomized) studies \citep{Sauerbrei14}.  The current
paper reflects the discussions within our STRATOS topic group (TG8),
and presents the first step toward a coherent approach to real-life applications of survival
analyses based on intensity (or `hazard') models. In particular, we discuss fundamental assumptions,
outline the basic steps necessary to ensure that the analysis
appropriately uses the data
at hand to address the substantive research question.
We also discuss some pitfalls and ways to avoid
them, point out some subtle complexities that may arise in applications, and suggest how the basic
methodology may be adapted or extended to address these additional
issues.

In many observational cohort studies, interest lies in
the occurrence of a particular \emph{event} among
subjects with a given condition (i.e. those `exposed') and among those
without the condition (`unexposed'), and the goal may be to compare
the pattern of event occurrence. In other settings factors of interest
may evolve over time as exposure changes with varying
treatments. Consider a register study of the
association between exposure to the drug
lithium and the incidence of dementia \citep{Kessing08}
(later referred to as the lithium and dementia study). In this study, the event is a
hospital diagnosis of dementia and the levels of exposure correspond to
different numbers of redeemed prescriptions (0, 1, 2, ...) of the drug
lithium. The lithium and dementia study will be used to set examples
throughout the first sections of the paper while other studies and data
sets will be used for illustrations in Section
\ref{sec:examples}. There, for example, we study the risk factors for all cause mortality in women with ovarian cancer and
 discuss more complex clinical cohorts and raise the issue of what can (and cannot) be
reliably estimated from the samples in the  studies of patients with non-alcoholic fatty liver disease
(NAFLD) and peripheral arterial disease (PAD).

\begin{figure}
	\includegraphics[width=15cm, height=6cm]{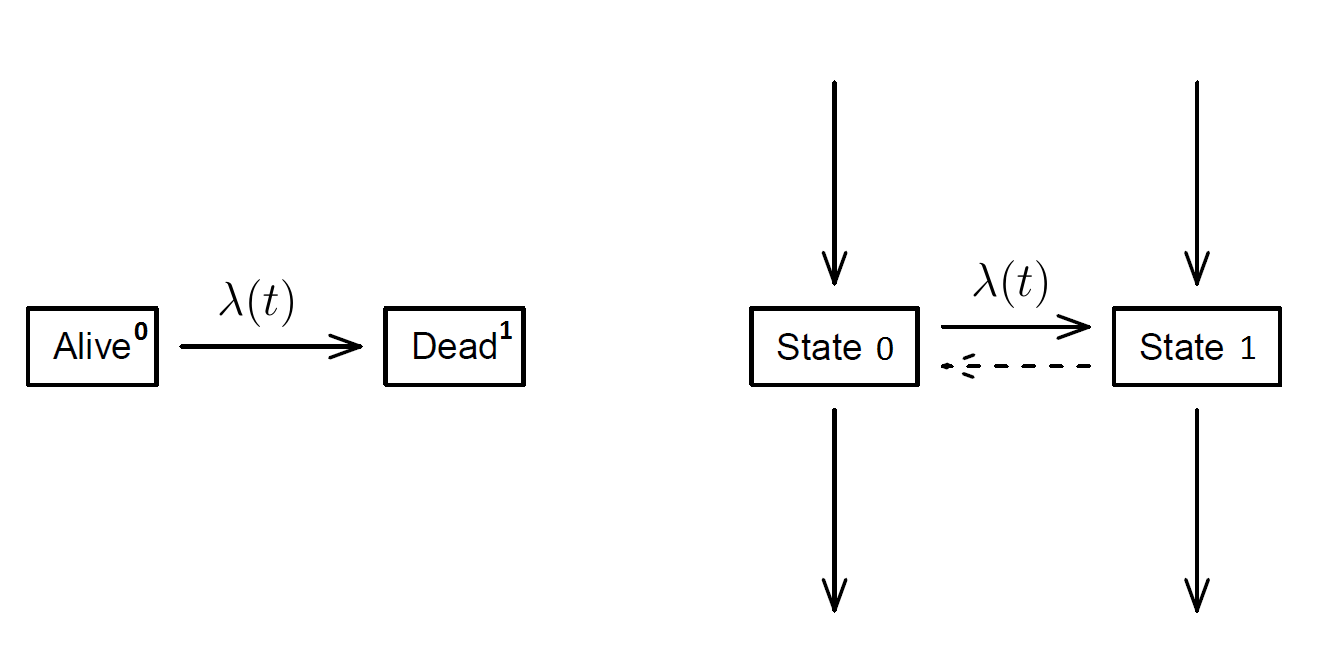}
	\caption{Representation of multi-state processes and transitions between states.  On the left
          is the simplest case of survival until death from any cause,
          on the right a
          more general situation where the transition of interest (from state 0 to state 1) is part of a larger multi-state model.}
	\label{fig:statediagram}
\end{figure}

The common situation for the examples is depicted in Figure
\ref{fig:statediagram}. In such settings there is a time axis usually measuring the time from the origin of the process or some other natural starting point that may be defined by the context. We denote this by $t$ and presume it is measured on a common scale for all individuals of interest.
The left-side graph refers to the simplest
situation, where death of any cause is the only event of interest
as in the ovarian cancer study. There the time origin would most naturally be the age of diagnosis with ovarian cancer.
The subjects
in the \emph{state} denoted as `0' are alive (with ovarian cancer) and, thereby,
\emph{at risk}  of experiencing the event  and the state corresponding to
this event is denoted `1'.
The right-side graph refers to a more general situation, where the
event of interest
may  be part of a larger system containing aspects that may or may not
be relevant for the study. In that case,
 the event may occur
via other states
than `0',  other events (transitions) may happen to subjects in state 0,
thereby preventing the occurrence of the event (competing
risks), and subjects may or may not return to state 0 after the occurrence
of the event (dashed line).   For subjects in state 0, there is a
probabilistic \emph{intensity}, say $\la(t)$ (or $\la_{01}(t)$), governing event occurrence and this intensity is often the primary target
for statistical modelling.

An important point to note in connection with such studies is that the
event of interest will typically not be observed for all
subjects. This incomplete information could be caused by different
mechanisms, including being event-free at end of follow-up (i.e.,
still in state 0 when the study ends or at an earlier time due to loss of
follow-up) or experiencing a competing event (i.e., leaving state 0 to
another state than 1). We will discuss the corresponding concepts of \emph{censoring} and \emph{competing risks} in detail in what follows.

We discuss survival analysis using intensity models
on data from cohort studies like those described in the examples above. We will
emphasize that there may be different types of scientific questions to
be addressed in such
observational cohort studies, and that the analysis
should be properly targeted to those questions.
Nonetheless there are a number of special features of survival data
arising from such studies of which investigators should be aware. In Section \ref{sec:basics}, we will discuss such features with examples
and give recommendations. We will also discuss potential pitfalls connected with
such analyses and how to avoid them.
In Section \ref{sec:hazmod} we will focus on models for the intensity $\la(t)$. We
will see that, for such an analysis, a more detailed description of
other (`competing') events that subjects may experience while being at
risk for the event of interest may not be needed.
However, an important point will be that even though such intensity
models may suffice for addressing some questions, one will have to go
beyond intensities to deal with other questions, including estimation
of the absolute risk of experiencing the event and also various \emph{causal} questions. This is the focus of Sections \ref{sec:beyond} and
\ref{sec:causal}. Section \ref{sec:examples} illustrates the relevant issues
and methods through analysis of the above examples, and the paper is concluded by a brief discussion in Section
\ref{sec:disc}.

\section{Getting the basics ready}
\label{sec:basics}

This section introduces the notation to be used throughout the paper
and defines the intensity. We also give a checklist of items that are
important to consider in observational time-to-event studies.

\subsection{Notation}
\label{sec:notation}
We assume that the following data can be available for subject $i$ in a sample of $n$ independent individuals, $i=1,...,n$:
\begin{itemize}
\item The follow-up time $T_i$, i.e., the time (relative to the chosen
  time origin) where the subject exited from the study.
\item The indicator variable $\delta_i$, indicating whether or not, at
  time $T_i$, the event of interest occurred ($\delta_i=1$ for event
  and 0 otherwise).
\item A time $V_i$ (relative to the chosen time origin where $V_i<T_i$)
  where the subject entered the study. Thus, if the subject was
  included already at the chosen time origin then $V_i=0$ , but $V_i>0$ (\emph{delayed
  entry}) is possible.
\item A vector of covariate values $Z_i(t)=(Z_{i1}(t),...,Z_{ip}(t))$, some of which
 may be  time-dependent but other may be time-fixed (baseline). 
\end{itemize}

Two common approaches exist for notation in survival data. A
traditional way to describe the survival data is the vector
$(T_i,V_i,\delta_i,Z_i(t))$, and this works well for time to all-cause death,
the simplest classic example of survival data (Figure \ref{fig:statediagram}, left panel). Alternatively,
one can view each subject as an evolution over time leading to the
counting process notation $(Y_i(t), N_i(t), Z_i(t))$.
This notation allows also more complex situations (as
implied from the right graph of Figure \ref{fig:statediagram}) to be
studied in a straightforward fashion and will be hence used throughout
the paper.
 Here, the process $Y_i(t)$ is equal to 1 while a person is known
 to be at risk
 and 0 otherwise, i.e. $Y_i(t)=I(V_i<t\leq T_i)$, and
 the process $N_i(t)$ denotes the number of events
 by time $t$, here simply $N_i(t)=I(T_i\leq t, \delta_i$=1).
Defining $dN_i(t)=N_i(t)-N_i(t-)$, observation of the event at time
$t$ can then be represented by $dN_i(t)=1$, i.e. the counting process for
$i$ counts +1 at that time.

\subsection{Preliminary concepts and issues}
\label{sec:check1}
The most important aspect of any analysis is to first think carefully about ($i$) what
question(s) we want to answer, and ($ii$) whether and how the data at hand
will be sufficient to answer them.
With respect to the latter ($ii$), important issues are the source of the
data, what population it
represents, what variables are relevant and which among these are available, and data completeness, both
with respect to inclusion of subjects and missing data for those that
are included.
We can then proceed with a more technical checklist:
\begin{itemize}
\item \textit{Time origin}: The follow-up time $T_i$ is measured from
  a meaningful starting point of the process ($t=0$), which should be unambiguously defined, comparable between
  individuals, and ideally clinically relevant.  Typical examples include age, time since diagnosis or time since treatment initiation. The choice of time origin should depend on the scientific questions.

In the lithium and dementia study, the time origin was defined as the date of the
start of the first lithium prescription for a given patient. Patients
who started lithium treatment after 1996 are included in the study. In the PAD study, the patients with
peripheral artery disease were identified at a visit to their
physician. While the onset of the symptoms may be observable,
it is impossible to know their times of the disease
onset.

The time origin does not always correspond to any `clinically
meaningful' date but to an administrative date of the `start of the
follow-up', so the study results must be interpreted accordingly.
In these cases age (as a major determinant of many health outcomes) may be the appropriate time axis, and it is increasingly used in population based studies investigating development of diseases.
If the underlying risks change systematically with e.g. time since diagnosis, then this will be the preferable time axis.

The definition of the time origin determines
  the primary time axis. In the lithium and dementia study, the time of
  interest is the time from treatment initiation to dementia
  onset. As the time origin is the time of lithium treatment initiation,
  all patients are at risk at time $t=0$, i.e. $V_i=0$ and $Y_i(0)=1$
  for all $i$. The study also addressed a second research question
  comparing the lithium treated patients to the general population and a
  sample from the general population is followed from 1st of January
  1995 onwards. In the general population, the time since lithium
  initiation of course cannot be defined and, thus, to answer the second research question,
   one should then use the   age as the time   \axis instead, for both the lithium treated and the general population subjects. This is an example of
  delayed entry for all individuals, everyone has $Y_i(0)=0$, since no
  one is included into the study at birth, the $Y_i(t)$ values
  switch to 1 at the age at which the individual was included in the
  study.
 Accounting for the delayed entry is necessary here to avoid so-called
 \emph{immortal time bias} \citep{suissa-AJE2007,anderson}, see Section
 \ref{sec:timebias}.

  To address the important question in the study
 \emph{multiple time axes} are needed -- both time since lithium treatment initiation
 and age. This is a common feature in epidemiological follow-up studies.
\item \textit{Inclusion criteria}: The inclusion criteria for
  individual $i$ must be met by the time that the patient is
  declared to enter the study, i.e., at time $V_i$ when $Y_i(t)$ first
  becomes 1.

   Say, we wished to analyse survival time of the general population in
   the lithium and dementia study: one cannot say that only
   individuals who are never treated with lithium throughout the
   study can be used for this analysis. While this information may be
   known later, at the date of data analysis, it was not known when
   the individuals were included.

   If the individuals who were treated with lithium at a certain
   point later in the study were excluded from the study, this would imply that patients
   with a shorter time to event (and hence less time to get lithium
   treatment) were more likely to be included and the survival of this
   general population group would be underestimated.

On the other hand, if 'ever treated' individuals from the general
population were included and were considered to be a part of the
patient group throughout the study, this would in turn imply the
overestimation of patients' survival. This is another example of
immortal time bias.  A correct way of analysis of
these data is to regard the exposure (lithium treatment) as a
time-varying covariate, we return to this in Section
\ref{sec:timebias}.

\item \textit{Event definition}: The time of event occurrence must be
  clearly defined. In the PAD study several events are of
  interest. The time of death or the time of a major cardiovascular
  event (stroke, infarction) are examples of well defined events with
  an exact date typically known. Other types of events, such as
  revascularisation procedure are slightly more subjective in nature, since this is an operation that has been
  scheduled following the doctor's evaluation of the patient's state
  of the disease. Judgement is therefore involved and moreover, the
  date when it took place depended not only on the stage of the disease but also the ability to schedule the procedure. Another example of an event with some ambiguity is the diagnosis of
  dementia. Here, one knows that the onset of dementia has happened between
  two consecutive visits, but is impossible to know when. This is
  referred to as interval-censoring. This could be further complicated
  if not all patient visits were scheduled with the same frequency, whereby
  some patients may be diagnosed earlier than others. In the lithium
  exposure study, these problems were avoided by \emph{defining} the event as
  the first hospitalization leading to a diagnosis of dementia (which may not coincide with the
  actual onset of the disease and must thus be interpreted
  accordingly).

The decision whether the event of interest has already occurred must be
known at the time when $N(t)$ switches to 1. A typical example occurs
with validated endpoints.  An endpoint such as diabetes might be
mis-coded, for instance, and investigators will often require `two
diagnoses at least 30 days apart' as proof.  An error occurs when
the date of diabetes is backdated to be that of the first instance and, therefore,
such an event definition would depend on something happening in the future.

\item \textit{Censoring}: The goal of the survival analysis is to
  estimate quantities relating to a complete, i.e. uncensored,
  population. A basic assumption in the estimation is that the
  information that a patient has \emph{not been censored} at a certain
  time point does not carry any information about his or her prognosis beyond
  that time point. We need this assumption since we shall regard the
  patients at risk  at a certain point $t$ ($Y_i(t)=1$) as a
  representative sub-sample of all the patients that would be at risk
  if there was no censoring. The assumption is referred to as
  \emph{independent censoring}.

     The assumption can be weakened to \emph{conditionally} independent
     censoring, i.e. independent censoring within a group of patients
     with a certain set of characteristics, defined by the covariates
     available in the study data base.
Administrative censoring, i.e. censoring at a certain calendar time
due to the end of study, is a common example of independent censoring,
as the censoring mechanism is not related to individual patient
prognosis. However, if the patient prognosis has improved through the
calendar time covered by the study, the patients diagnosed later have
a better prognosis and are also censored earlier, so the independent
censoring assumption is not met. On the other hand, the censoring pattern is
conditionally independent given the period of diagnosis, so inclusion of this covariate
in the model will avoid potential bias (provided  the model is correct).
It is, thus, important to consider what causes censoring in any given follow-up study.
   Is it mostly administrative censoring, i.e., being event-free at end of planned follow-up, or are there
   drop-outs? While the former can often be taken to be independent, more suspicion should be exercised
   for the latter and, ideally, it should be noted in the data set \emph{why} any given subject was censored.

    In some studies one considers  deaths due to causes
    unrelated to the disease in question as censoring, e.g. regarding the
    patients who die from non-CV causes in the PAD study (causes that
    were not of the main interest in this study) to be equivalent to
    those who were administratively censored. This is not a formal
    violation of the definition of independent censoring, rather it is
    inconsistent
    with the definition of our population of interest. This is because
    the competing
    risk (death from non-CV causes) is present also in the complete
    population and we usually are not interested in the population where this
    risk would be eliminated.
      We address this situation in the Competing risks
     section (Section \ref{sec:comprisk}).
\end{itemize}

\subsection{The intensity}
\label{sec:intensity}
We now return to the concept of the \emph{intensity} function.
Once it has been established who is at risk, what is the event and how should subjects should be aligned over time (i.e. how time $t$ is defined),
one can define the \emph{intensity} for subject $i$ as:

\begin{eqnarray}
\la_i(t)&\approx& P(\mbox{event in }(t,t+dt)\mid \mbox{past at time }t-)/dt \label{eq:int}\\
  &=& P(dN_i(t)=1\mid H_i(t-))/dt,\nonumber
\end{eqnarray}
where, $H_i(t-)=(N_i(s), Y_i(s), Z_i(s), s<t)$ summarizes the past
information for the subject $i$ that is available just before time
$t$.
(More formally: $\la_i(t)=\lim_{\Delta t\rightarrow 0}P(dN_i(t)=1\mid
H_i(t-))/\Delta t$.)

For \emph{survival data} involving only states 0 and 1 as in the left
panel of Figure \ref{fig:statediagram},
   the \emph{hazard function} is given by:
\begin{equation}
\al(t)=-\frac{d\log S(t)}{dt},
\label{eq:hazard}
\end{equation}
where the \emph{survival} function $S(t)$ is $P(T^*>t)$ and
$T^*$ is the uncensored -- and incompletely observed -- time
     to death. When no time-dependent covariates are considered in
     $H_i(t)$, the intensity in (\ref{eq:int}) is simply given by the
     hazard in (\ref{eq:hazard}) and, for that reason, we will use the
     terms \emph{intensity} and \emph{hazard} interchangeably in this paper.
While the hazard function for survival data (as seen in equation
(\ref{eq:hazard})) is in one-to-one correspondence with $S(t)$, and
thereby with the cumulative risk $F(t)=1-S(t)$, there may more
generally be other
events competing with the event of interest in which case $\al(t)$ is
a transition hazard (or cause-specific hazard) from state 0 to state
1 (and it is some times denoted $\al_{01}(t)$ to emphasize this). In
that case, a one-to-one correspondence between the
single cause-specific hazard and the absolute probabilities no longer
exists, see e.g. \cite{PKA-IJE}.

  There are several reasons why the intensity function plays a central
  role in survival analysis and analysis of cohort studies.
   \begin{itemize}
   \item The idea is that subjects are followed over time and, at each
     time $t$ where the subject is still observed to be at risk, it is
     asked, given the information that is available so far for
     the subject, what is then the probability per time unit that the
     subject experiences the event in the next little time interval
     from $t$ to $t+dt$? Thus, the intensity gives a dynamical
     description of how events occur over time and, in this
     description, all aspects of the \emph{past} observed for the
     subject up to time $t$ may be taken into account, such as (time-dependent) covariates and, possibly, previous events.
   \item  For survival data, the survival probability $S(t)=P(T^*>t)$ cannot be
     estimated in a
   straightforward way as a simple proportion due to censoring. The
   hazard function, however, can still be studied  since it relies exclusively on the information on the
   individuals still at risk
   (assuming the independent censoring assumption to
   hold).
   \item One of the greatest strengths of describing the data via the
     intensity is that the past of a subject could include not only
     the \emph{baseline covariates}, i.e. information available at time
     of entry into the study but also information
     contained in \emph{time-dependent} covariates that are updated during follow-up. Time-dependent
     covariates have a very natural connection to clinical practice:
     when assessing a long-term patient, a physician will naturally
     use the most recent measurements in addition to those collected a long
     time ago at their initial visit. Time-dependent covariates thus
     allow accounting for the changes over time in the relevant variables (patient characteristics, exposures, treatment), which may alter the
     intensity.

     In the lithium and dementia study, the exposure itself (number of
     lithium prescriptions redeemed until the current time) is time-dependent and, when
     including the unexposed control group of non-treated subjects,
     people will change status from being unexposed to belonging to the
     exposed group at the time of their first lithium prescription.

    Unfortunately, misunderstandings and errors in creating
    time-dependent covariates are one of the most common sources of
    immortal time bias, we thus pay this issue special attention in
   Section \ref{sec:timebias}. Time-dependent covariates also create issues
    with model prediction, which will be discussed in Section
    \ref{sec:Prediction}.
     \item As the intensity is a dynamic description of the data generating process over time,
    delayed entry is naturally taken into account.
\item    Most often, the hazard depends on patient characteristics, so that
\emph{covariates} need to be taken into account when analyzing data
from cohort studies. In this context, as discussed in Section \ref{sec:check1} above, an advantage of using a hazard regression  model is that it
`corrects for non-independent censoring' in the sense that regression
coefficients are estimated consistently even if censoring depends on
covariates, as long as these covariates are included as predictors in the hazard
model (e.g., \cite{ABGK-book}, Section III.2).
\end{itemize}
  We argue that the intensity may be of interest in its own right and that it,
  therefore, could be an obvious target for an analysis of the cohort
  data. However, as previously mentioned, there are important
  scientific questions for the answer of which the intensity will be
  insufficient. We will return to that in Sections \ref{sec:beyond} and \ref{sec:causal}.

\section{Hazard models}
\label{sec:hazmod}

In order to make a \emph{model} for the intensity one needs to specify
how it depends on time $t$ and on the available information in
$H_i(t)$, more specifically: how it depends on the covariates.

Before specifying a hazard model, descriptive analysis should be conducted to explore the data.
The \emph{Kaplan-Meier estimator} for the whole cohort, or for
sub-groups, will provide useful insights in the case of a single (possibly composite) terminal event, see the left panel of Figure \ref{fig:statediagram}.
However, this will not be the case
in, e.g. the lithium and dementia study where, obviously, all-cause
mortality is a competing risk for dementia, the event of interest. To
describe the \emph{hazard}, one could divide the time variable of
interest into suitable intervals (e.g., yearly intervals) and
calculate the \emph{incidence} (or `occurrence-exposure') rate for each interval by
dividing the total number of events of interest observed in the
interval by the total time at risk in the interval. This corresponds
to an assumption of a piecewise constant hazard function, an
assumption that is often a reasonable approximation and which is used
for \emph{Poisson regression} which will be discussed further
below.

To estimate the hazard \emph{non-parametrically}
requires some sort of smoothing. What may be estimated in a simple
non-parametric fashion for a defined population of subjects is the
\emph{cumulative hazard} $\Al(t)=\int_0^t \al(u)du$. This may be done
using the \emph{Nelson-Aalen} estimator
\begin{equation}
\widehat{\Al}(t)=\int_0^t\frac{dN(u)}{Y(u)}=\sum_{i:T_i\leq t}\frac{ \delta_i}{Y(T_i)}
\label{eq:NAa}
\end{equation}
where $Y(t)=\sum_i Y_i(t)$ and $N(t)=\sum_i N_i(t)$. This is an
increasing step function with steps at each observed event time, the
step size at $t$ being inversely proportional to the number $Y(t)$ at risk, Figure \ref{fig:nafld2_na} shows an example. Pointwise confidence
limits may be added (e.g., \cite{ABGK-book}, Chapter IV). The
approximate `local slope' of $\widehat{\Al}$ at $t$ reflects the hazard at
that time. However, the value of the cumulative hazard,
does, in general, not have a simple interpretation (an exception is in studies of \emph{recurrent} events, a topic
beyond the scope of this paper \citep{cook-lawless-book2007}.

\subsection{Proportional hazards models}
\label{sec:Cox}
Most applications of intensity models for cohort studies aim at relating the rate of event
occurrence to (time-fixed or time-dependent) \emph{covariates}. The complexity of such a hazard
model will both depend on what is the scientific question that it is meant address and on
what information is available. Here, it is important to notice that hazard models, as any other statistical model, can typically not be
expected to be `correct' in any strict sense but still they may be sufficiently flexible to give
a sensible answer to the question raised.

Survival analysis and, thereby, also analysis of cohort studies is dominated by the Cox model. In his breakthrough
paper, \cite{Cox72} introduced the maximum partial
likelihood as a method to estimate the regression coefficients in the proportional
hazards (PH) model. The general definition of the proportional hazards model is
\begin{eqnarray}
\al_i(t)=\al(t|Z_i(t))=\al_0 (t)\exp(Z_i(t)^\top\beta). \label{eq:PHmodel}
\end{eqnarray}
In the Cox PH model, the \emph{baseline hazard} $\al_0(t)$ (i.e., the hazard for individuals with all covariates equal to 0) is left
unspecified, other alternatives for the PH model are considered at the
end of this section. The name `proportional hazards' refers to the,
possibly strong, assumption that the ratio of the hazards corresponding to two
different values of $Z_i(t)$ is the same  for all times $t$. This constant hazard ratio is $\exp(\beta)$. In this
model, the  regression parameter(s) $\beta$ are estimated by
maximizing the log partial likelihood
\begin{equation}
pl(\beta) = \sum _{i=1}^n \int_0 ^\infty \log \left(
\frac{\exp(Z_i(t)^\top \beta)}{\sum_{j} Y_j(t) \exp(Z_j(u)^\top
  \beta)} \right)dN_i(t).
\label{eq:CoxPL}
\end{equation}
Once the covariate effects $\beta$ have been estimated, the cumulative baseline hazard $\Al_0(t)=\int_0^t \al_0(u)du$ can be estimated  by
the Breslow estimator \citep{Breslow74}
\begin{equation}
\widehat \Al_0 (t |\widehat \beta) =\int_0^t { { d{N} (u)} \over {
    \sum_i Y_i(u) \exp(Z_i(u)^\top \widehat{\beta})}}.
\label{eq:breslow}
\end{equation}
With no covariates ($\beta=0$) the Breslow estimator is simply the
Nelson-Aalen estimator (\ref{eq:NAa}). The partial likelihood may be seen as a profile likelihood resulting from eliminating
the baseline hazard from a joint likelihood including both $\beta$ and $\al_0(t)$. It is important to notice that this joint likelihood
is valid both for all-cause survival data (left panel of Figure \ref{fig:statediagram}) and for the more general
situation depicted in the right panel of that figure. This is because the likelihood for the whole system
\emph{factorizes} into a number of factors, each depending on a
separate transition in the model \citep{Kalbfleisch02, ABGK-book}.
Also, the Cox partial likelihood (\ref{eq:CoxPL}) enjoys the properties of standard likelihood functions such that
standard errors and test statistics may be obtained in the `usual way'
\citep{Cox-75, andersen-gill-AS1982}. We will use these features
in the Examples section \ref{sec:examples}.

If more time axes are relevant in a given study then, using the Cox model, one of these must be selected as the baseline time axis $t$.
Other time axes (e.g., current age if $t$ is time since disease onset)
can then be included as time-dependent covariates. This type of
time-dependent covariates is said to be \emph{external} (or \emph{exogenous}) because they `exist' whether or not the subject is still
under observation. On the other hand, time-dependent covariates such
as current blood pressure or current cholesterol level (that can only
be ascertained for subjects still under observation) are \emph{internal} (or \emph{endogenous}).

The completely unspecified baseline in the Cox model makes it quite flexible, however,
a limitation of this non-parametric model component is that it
only allows direct estimation of the \emph{cumulative} baseline
hazard $\Al_0(t)$, but fails to produce an estimate of the hazard
$\al_0(t)$ itself. To obtain an estimate of $\al_0(t)$, some smoothing would be required.

The alternative to letting the baseline hazard remain unspecified in
the model is to fit a fully parametric proportional hazards model,
some of the common options are:
\begin{itemize}
	\item The simplest (and most restrictive) option is to assume a \emph{constant}
          baseline hazard corresponding to an exponential distribution of $T^*$ in the case of
          all-cause survival data.	
	\item A useful extension of the model above
is the \emph{piecewise exponential} model that divides the
          time-range into intervals on which the baseline hazard is
          constant. Here, cut-points
          $0=s_0<s_1<\dots<s_{K-1}<s_K=\infty$ for the time axis are
          selected and it is assumed that $\lambda_0(t)=\lambda_{j0}$
          when $s_{j-1}\leq t<s_j$.
This is often referred to as the \emph{Poisson}
            (or piecewise exponential) regression model for survival data and it is frequently
          used in epidemiological studies. It tends to produce results
          that are very close to those obtained using the Cox
          model. An advantage is that the baseline hazard is fully
          parametric and yet flexible. Further advantages are that a
          potentially large (e.g., registry-based) data set can be
          pre-processed into tables of \emph{event counts} and \emph{person-time at risk} according to the chosen time
          intervals and to (discrete-valued) covariates and, furthermore, that
          \emph{multiple time axes} are very easily handled  by
          splitting event counts and person-years at risk
          simultaneously according to all time axes \citep{CH-book}.
           Drawbacks of the piecewise exponential model include the fact
          that the intervals must be selected and that this choice to
          some extent may affect the detailed results and, furthermore, that it does not produce a \emph{smooth}
          hazard function.
\item           A smooth extension of the constant baseline hazard model is
the Weibull model, where $\al_0(t)=\la \gamma(\la
          t)^{\gamma-1}$. The extra parameter $\gamma$ allows some
          flexibility, but assumes a monotone baseline hazard
          function and the model is not flexible around $t=0$.
To allow greater flexibility and obtain a smooth baseline hazard one may use flexible parametric models for
          $\al_0(t)$, e.g., via splines in combination with penalized
          likelihood \citep{joly-etal}.
\end{itemize}

\subsection{Alternatives to proportional hazards models}
\label{sec:alternat}
The PH assumption is strong and may often not fit the data well for the entire time range
studied.
An extension of the Cox model, to relax the PH
assumption, is to allow the covariates $Z(t)$ to have \emph{time-varying}
effects, i.e., assume that the hazard is given by
\begin{equation}
\al(t|Z_i(t))=\al_0(t)\exp(Z_i(t)^\top\beta (t)).
\label{eq:timevarcoeff}
\end{equation}
Here, explicit
   interactions between covariates and functions of time may be
   introduced, e.g., by defining a model with $\beta(t)=\sum _j
  \gamma_jf_j(t)$ (for a set of pre-specified functions $f_j(t)$
  containing $f_0(t) \equiv 1 $) for each component of $Z$. The simplest example is
  to split time into two intervals (splitting at time $\tau$, say) and assume proportional hazards within each. This
  corresponds to choosing $f_0(t)=1\mbox{ and }f_1(t)=I(t\geq
  \tau)$. Alternatives include the use of splines \citep{royston}. Some care
  is needed here since the number of parameters in such models with
  time-varying effects can become quite large  and there is a danger
  of overfitting. In the situation where the PH assumption needs to be relaxed
  for a single categorical covariate, the \emph{stratified Cox model} is useful.
  In this model, each level of that covariate has its own baseline hazard which is
  not further specified, i.e.
  \begin{equation}
\al(t|Z_i(t))=\al_{0s}(t)\exp(Z_i(t)^\top\beta),
\label{eq:stratCox}
\end{equation}
when subject $i$ belongs to stratum $s$.

An alternative to the multiplicative Cox model is the \emph{additive
  hazards} (or \emph{Aalen}) \emph{model} \citep{Aalen89}
\begin{equation}
  \al(t|Z_i(t))=\al_0 (t)+ Z_i(t)^\top\beta(t).
\label{eq:Aalenmodel}
\end{equation}
In this model, both  the baseline hazard $\al_0(t)$ and the
regression functions $\beta(t)$ are completely unspecified
(like the baseline hazard for the Cox model) and their cumulatives
$\int_0^t \al_0(u)du, \quad\int_0^t \beta(u)du$ can be estimated using a least squares technique.
Versions of the model where some or all $\beta(t)$ are time-constant
are also available \citep{Scheike07}.
A drawback of the model is that the estimated hazard can become negative while an advantage is that it is very flexible \citep{aalen-book2008}.

A completely different approach  is given by the accelerated failure
time (AFT) model where the covariates are assumed to extend or shorten the survival time by
a constant time ratio $\exp(\beta)$ \citep[e.g.][Ch. 7]{Kalbfleisch02}
$$  S(t|Z) = S_0(\exp(-Z^\top \beta)t),  $$
or  equivalently:
$$ \ln(T^*)=Z^\top\beta + \epsilon .  $$
The model is a viable alternative to the PH model and although
one could derive the hazard function for this model, it does not naturally fall under the
heading of `hazard models'. Also, it is mostly used for survival data (Figure \ref{fig:statediagram}, left part) and less so in the general situation of Figure \ref{fig:statediagram}, right part, and it will not be considered further in this paper.  Discussion of pros and cons of PH and AFT can be found in \cite{wei92} and \cite{Keiding97}.

\subsection{A checklist when fitting the Cox model}
\label{sec:checkCox}
We next propose a checklist for the Cox model. Most of the items below
are relevant also for other hazard regression models.
We list the issues that one should be careful about both before fitting a model
and after having performed an
analysis. We add tests and approaches that can be helpful in
understanding the sources of the problems and evaluating their
extent. Note, however, that these checks are not conclusive, they serve
only as an aid in thinking about the issues.

Before fitting a model the following items should be considered:
\begin{itemize}
  \item	\textit{Checks on the covariates} to be included in the model:
    for continuous covariates
    examine the distribution,  check for
    extreme data (leverage) points, make histograms.
    For categorical covariates, the frequencies of the categories should be
          reported and also the choice of the reference
          categories.
   \item       \textit{Check dates.} A trivial, but often relevant warning is that
  survival data often contain a series of dates, that may come in
  varying formats and are prone to typing mistakes. The fact that the
  dates follow each other in the proper sequence should thus be
  carefully checked.
   \item \textit{Investigate censoring.} As mentioned in the previous checklist, it is first  of all
   important to think about what causes censoring. Next,
   plotting a `survival curve' estimating $C(t)=P(\mbox{no censoring before }t)$ (or its complement
   $1-C(t)$)  could be
          done to give an impression of the proportion censored in time. Here, censoring is the `event' and a failure is a `censoring event' that prevents observation of the `event of interest'. Also, a Cox regression model
   with `censoring' as event can help to check whether the censoring
   depends on any of the covariates under consideration. If there are
   some variables that one may or may not include in the model (maybe
   they are not crucial for the question asked) then they should be
   included if they affect the censoring, since in this way the independent
   censoring assumption is relaxed to conditional independence, as discussed in Section \ref{sec:check1}.

An important feature of hazard models is that they can be used exactly
as described in Section \ref{sec:basics} by \emph{formally censoring for the competing
  events (including all-cause death)}. This is not a violation
of the independent censoring assumption, the point being, as mentioned above, that
the joint likelihood function for both the event of interest and the
competing events \emph{factorizes} and the factor corresponding to the
intensity for the event of interest has the same form as it would have
had if competing events were regarded as
censoring events \citep{Kalbfleisch02}.  In such situations one should
carefully consider if
the (cause-specific) hazard for the event of interest properly answers the scientific question
or whether one needs to go beyond this hazard model (see Sections
\ref{sec:beyond} and \ref{sec:causal}).
\item	\textit{Time-dependent covariates.} When defining the
          model in (\ref{eq:PHmodel}), we assume that the $Z_i(t)$ are
          continuously measured and, thus, available at all times $t$,
          for which subject $i$ is at risk.

  A feature of the partial likelihood estimation method for the Cox
  model is that the values of  time-dependent covariates are
  needed for everyone at risk at all the event times, cf. equation (\ref{eq:CoxPL}). Some extrapolation or other ways of predicting the value of a time-dependent
  covariate at event times based on \emph{past} observations $(Z(s), s\leq t)$
  may be needed \citep{bycott-taylor}. In practice, most recently observed values of $Z(t)$ are typically carried forward until
    the next value is observed. However, such
    last-value-carried-forward approach can induce some bias toward
    the null if the current hazard depends truly on the current
    (unknown) covariate value \citep{andersen-BIOSTATISTICS2003,
      de-bruijne-SIM2001}.
A more advanced approach for internal time-dependent covariates which
are not measured at all times uses joint longitudinal-survival models
\citep{wulfsohn-tsiatis, tsiatis-davidian} to obtain
estimates of $\beta$, and also allows the possibility that the
observed $Z(t)$ is measured with error.
     Note that, for external time-dependent covariates, extrapolations
     are sometimes not needed since, e.g. current age can be
     calculated based on age at baseline.

Covariates that change shortly before the endpoint should be
viewed with particular suspicion.
A common example is a change in medication in the last 1
or 2 weeks before death; such changes often occur when a patient
enters terminal hospice care for instance.  The most serious examples
of such `anticipation'
involve \emph{reverse causality bias} where a change in $Z(t)$ occurs
\emph{because of} early symptoms of the event of interest
\citep{horwitz-feinstein}. In some applications it may be therefore
more plausible that the current hazard depends on the past rather than
most recent value(s) of a time-dependent covariate implying either
lagged or cumulative effects that would require more complex modelling
\citep{gasparrini, sylvestre}.
\end{itemize}
After having fitted a Cox model one should consider:
\begin{itemize}
  \item	\textit{Check proportional hazards and the functional form}.
   Two basic assumptions of the model are that the coefficients
   $\beta$ are time-fixed (PH assumption) and that the covariate effect
   is linear on the log hazard. Checking the PH assumption has
   developed into a large
`industry' within survival analysis and giving a comprehensive review
is beyond the scope of the present paper. Among the many methods proposed
(some of which will be illustrated in the Examples section \ref{sec:examples})
we mention those based on Schoenfeld and martingale residuals
\citep{therneau00, Lin93}, graphical methods such as plots of the
cumulative hazard \citep{ABGK-book}, or through estimates of $\beta(t)$
in a time-varying coefficient Cox model \citep{Scheike07}.
For relaxing the linearity assumption,
  one may wish to use simple transformations like the logarithm or the
  square root
  or, alternatively, flexible modeling using, e.g., splines \citep{royston}.
    For continuous
   covariates, functional form (i.e., non-linear effects) should ideally be
   investigated jointly with assessing possible violation of the PH
   hypothesis (i.e, their `time-varying effects'). Indeed, a failure
   to account for a time-varying effect may induce a `spurious
   evidence' of non-linearity and vice versa
   \citep[e.g.][]{Abrahamowicz07,Wynant14}.

Another question is what to do if model assumptions seem to be
violated. Here, the answer must depend on what are the consequences of
the model violation. In a classical epidemiological
`exposure-confounder' situation, if the assumptions do not hold for
some of the \emph{confounders}, one may wish to perform a sensitivity
analysis. Specifically, to relax the PH
   assumption, one can introduce time-varying effects $\beta(t)$ in
   the model (see (\ref{eq:timevarcoeff})) or use a stratified model and if the results for the exposure in the
   sensitivity analysis do not change materially, the assumption may not be
   problematic.
If, on the other hand, the assumptions do not hold for the exposure, one
should carefully think about the study question and then employ
extensions of the basic model if needed. In such cases modelling the time-varying hazard ratio may yield important insights into the role of a given exposure or risk factor. Note that violation of the PH assumption may be some times
induced by a failure to include in the Cox model a strong predictor of
the outcome \citep{schmoor, bretagnolle}.
	\item \textit{Reporting.}
Users of the Cox model often report the regression coefficients,
but not the baseline hazard. This means that measures like absolute risk
cannot be retrospectively obtained from published reports.
This is insufficient  because the regression parameters
        figuring in the partial likelihood  only give
        information about the hazard ratios and the relevance and importance of the
        hazard ratios at any follow-up time depends on the concurrent values of the
        baseline hazard.

        The discrete nature of the estimated baseline
        hazards in the Cox model makes it hard to compare the
        hazards. For survival data, the estimated survival probability $\widehat
        S(t|Z)$ can be used to quantify the effects of $Z$. This is only possible if $Z(t)$ is a time-fixed or
       an external
        covariate (see Section \ref{sec:beyond}). Predicted survival curves for a population may be
        calculated by averaging over the observed covariate
        distribution (using the $g-$formula, see Section \ref{sec:causal}, equation (\ref{eq:g-formula})).
	\item\textit{Interpretation} Three phenomena hamper the
          interpretation of the results (hazard ratios) of a Cox model:
	\begin{itemize}
	\item \textit{Noncollapsibility.} It is frequently seen that
          the effects,
          $\beta$, in a Cox model gradually decay with time toward 0. This
          happens even if the
          true effect (i.e., given all relevant covariates) is
          perfectly constant over time if a covariate
          with an effect on the hazard is omitted, even if that
          covariate is completely independent of the other
          covariates. Thus, if the correct model is $\al(t)=\al_0(t)\exp(\beta_1Z_1+\beta_2Z_2)$
          then a reduced model $\al(t)=\widetilde{\al}_0(t)\exp(\widetilde{\beta} Z_1)$ cannot hold even if $Z_1$ and $Z_2$
          are independent, so that $Z_2$ is not considered a
          confounder for $Z_1$ \citep{struthers1986}.
            The non-collapsibility suggests that proportional hazards can
          only be seen as a working hypothesis allowing a simple
          structure. It can be noted that the logistic regression model for a binary response variable
          suffers from the same problem, while the additive hazards model does not.
	\item \textit{Competing risks.} The function obtained
          from the Cox model (or any other hazard model) using the
          formula $F(t|Z)=1-\exp(-{\Al}((t|Z))$)
          can only be interpreted as the risk of failure up to $t$
          if there are \emph{no other causes of death}. If dying from
          other causes (competing risks) is handled as censoring, the
          resulting function will over-estimate the probability of the
          event of interest \citep{PKA-IJE}. This must be represented by a cumulative
          incidence function instead, see Section \ref{sec:beyond}.
	\item \textit{Lack of causal interpretation.} Suppose the estimated hazard
          ratio for a treatment variable changes over time in such a way that,
          before some time point $\tau$, it is less than 1  (suggesting
          a beneficial effect) and after $\tau$ it is equal to 1, i.e.:
          $$\al(t)=\al_0(t)\exp(\beta_1ZI(t<\tau)+\beta_2ZI(t\geq \tau))$$
          with $\beta_1<0, \beta_2=0$. Even though this may be a correct model for the data
          it would be incorrect to assert that `treatment only has an impact
          before time $\tau$'. This is because the hazard does not provide a
          `causal contrast' \citep{hernan-EPIDEMIOLOGY2010,
            aalen-LiDA2015, torben-etal}. We will also elaborate
          on this point in Section \ref{sec:causal}.
\end{itemize}
\end{itemize}

\subsection{Time-dependent covariates and immortal time bias}
\label{sec:timebias}

In the process of creating the data set it is all too easy,
unfortunately,  to ignore the fact that a covariate is time-dependent
and treat it as time-fixed. This is a common source of `immortal
time bias'  and may be the single most prevalent reason for invalid
survival analyses in the literature.

  It is crucial that only information reflecting covariate values observed before time $t$
  (i.e, from the `past' at $t$) is
  used to define the value of a variable $Z(t)$ at time $t$.  Thus,
  even though later information pertaining to changes that occurred \emph{after} $t$ may be available to the investigator
  at the time of analysis, only information for a given subject that
  reflects changes that occurred before time $t$ can be included as
  part of that subject's past at time $t$. We give a couple of the most
  common examples with invalid $Z(t)$, but these do not nearly exhaust
  the possibilities.
  \begin{itemize}
    \item A common example is to group patients at time $t=0$
      according to the use of drugs or treatments at any time during the follow-up even if many might have started their use only at some time after time $t=0$ (`ever treated' vs `never treated'). \\
      E.g., in the lithium and dementia study, as mentioned above, one cannot
         say that only individuals that are never treated with lithium
         throughout the study serve as a control and thus create a
         time-fixed covariate `exposure'. While the information that
         an individual has not been treated throughout our follow-up
         is known at the time of data analysis, it could not be known when
         the patients entered the cohort. The exposure should, thus, be coded
         as a time-dependent covariate that starts with 0 for all
         individuals that were sampled from the general population,
         but may switch to 1 at a subject specific time $t_i$ if subject $i$ did start lithium treatment at that time. The
         alternative, i.e. to treat all individuals sampled from the
         general population as un-exposed regardless of what happens
         to them later will bias the comparison in the direction
         of ‘protective effect’ of lithium exposure, as explained in Section \ref{sec:check1} above.\\
         Another classical example of the same problem is the Stanford
         heart transplant data \citep{crowley-hu}, which included
         patients who were eligible for heart transplant. The event of
         interest was death and the focus was how the survival of
         transplanted subjects compares to that of not
         transplanted. When the data were first analysed \citep{clark}, the patients were
         divided into two groups (`ever transplanted' vs. `never
         transplanted'), the group membership was wrongly represented as
         a time-fixed covariate. As it turned out later with correct
         analysis, original results which suggested that the transplant is beneficial
 were solely due to the immortal time bias.
    \item  A similar situation occurs when studying a trait that
      develops in time (e.g., with time patients develop side effects
      or, with time, patients may respond to chemotherapy). Here,
      the  value of their covariate starts as 0 and may become equal
      to 1 later. This automatically implies that individuals must
      survive at least some time to develop the trait, the early
      deaths are hence more likely to occur in patients without the
      trait. Considering the value of the covariate as time-fixed
      and wrongly coding it as 1 already at the start (ever developed a
      trait or ever had side effects), will mis-attribute the portion of event-free survival time from the 'unexposed' (no trait yet) to the 'exposed' group and, thus, underestimate the hazard ratio associated with having the trait.
    \item A further example is to model the total dose received during the entire follow-up period as a
      time-fixed variable.  \cite{Redmond83} investigated
      a claim of \cite{Bonadonna81}  that
      disease-free survival improved with increased total amount of drug received, and found it
     to be entirely due to immortal time bias because the patients who died
     early could not have accumulated high doses.
    The false result is not benign, since it would encourage providers to
    continue full dose treatment in the face of dose-limiting toxicities, leading
    to increased morbidity, suffering, and possibly even death.
  \end{itemize}
In all of the above cases, creation of a well-defined time-dependent covariate
 where $Z(t)$ does not depend on any of
$Z(s), N(s)$, or $Y(s)$ for any $s > t$ repairs the bias.

\section{Prediction using hazard models}
\label{sec:beyond}

Even though the intensity discussed in previous sections provides a
useful framework for statistical modelling it may be hard to explain the model results
to the general public. Communication is usually easier in terms of
the \emph{absolute risk}, i.e., the
probability of the event occurring in some interval or, more generally, the probability of being in a certain \emph{state} by time $t$. For estimation of
the absolute risk it now becomes crucial to consider if
other (competing) events may occur, thereby preventing the event of interest
from happening, see Figure \ref{fig:statediagram}. In general, \emph{all} transition hazards out of the initial state, 0,
are needed to estimate absolute risks.

\subsection{Prediction in the absence of competing risks}
\label{sec:Prediction}

In the case of no competing risks, there is only one hazard function in the model and the absolute risk for the interval from
0 to $t$ is obtained directly from that hazard:

\begin{equation}
F(t|Z)=1-S(t|Z)=1-\exp(-\Al(t|Z))
\label{eq:absrisk}
\end{equation}
provided there are no  time-dependent covariates. Absolute risk is
often used to describe survival or recurrence-free survival in clinical
cohorts of patients treated for cancer or other life-threatening
diseases.

\subsubsection*{Prediction from $t=0$ onwards}
This is relevant when the time origin is well-defined. In clinical
cohorts it can
be the time of diagnosis or start of treatment. In population cohorts
it can be a fixed value of age. The
prediction is given by the  survival probability $S(t|Z)$ where $Z$ contains the
relevant information available at $t=0$.

\begin{itemize}
	\item \textit{Using the Cox model:} The PH assumption is often
          not satisfied over the whole time range. That could be fixed
          by introducing time-varying effects of the covariates (or by finding another model that fits
          the data better, e.g. an additive hazards model). If
          the focus is on one particular value $t_s$ (e.g., the 5-year
          survival in cancer), a surprisingly robust estimator of
          $S(t_s|Z)$ can often be
          obtained by applying administrative censoring at $t_s$ and
          using a simple Cox model with the effects $\beta$ fixed in
          time provided $t_s$ is not too large \citep{hans-stopped}.
	\item \textit{Direct modeling:} If there were no censoring
          before $t_s$, the survival probability $S(t_s|Z)$ could be directly
          estimated by models for the binary outcome $I(T^* >
          t_s)$. There is a choice of link function: probit, logit
          and complementary log-log. The latter measures the effect on
          the same scale as the PH model. Models can be fitted using full
          maximum likelihood or estimating equations
          approaches. Censoring before $t_s$ can be handled by modeling the censoring distribution and using
          inverse probability of censoring weighting (IPCW) or by
          using pseudo-observations based on jack-knifing \citep{PKA-maja}.
\end{itemize}
\subsubsection*{Dynamic prediction}
Predictions made at $t=0$ need to be updated later on for those
individuals that are still alive and at risk for the events of
interest. First of all, the survival  probabilities have to be replaced
by the conditional probability $P(T^* > t | T^* \ge t_{pred}, Z(t_{pred}))$, where
$t_{pred}$ is the time from which a new prediction  is wanted. If the model for
the hazard is perfect, the conditional probability can directly be
computed from the hazard using
$\widehat{\Al}(t|Z(t_{pred}))-\widehat{\Al}(t_{pred}|Z(t_{pred}))$  for $t \ge
t_{pred}$. However,
it may be hard to make models that are valid over the whole time
range. Therefore, an alternative is to develop a new model using the data of
the individuals still alive at $t_{pred}$. If there is a fixed
prediction window $t_s$, the conditional survival $P(T^* > t_{pred}+t_s|T^* >
t_{pred},Z(t_{pred}))$ can be estimated robustly by the methods discussed above.
Prediction later on is known as \textit{dynamic prediction} and the
approach of building new models using the individuals at risk at
$t_{pred}$ is known as \textit{landmarking} \citep{hans-sjs}.
This is also of interest more generally when
there is \textit{delayed entry}, i.e. individuals entering the cohort
at $V>0$. In this case, the hazard can be hard to estimate around $t=0
$, since only few individuals may be at risk early on. Hence, predictions are hard to make at $t=0$, but conditional
survival probabilities could be estimated reliably later in the
follow-up. That might be particularly relevant for analyses that use
age as the time axis.
\subsubsection*{Prediction exploiting  time-dependent covariates}
Dynamic predictions using landmarking can thus be used when the PH assumption does not give a reasonable
description over the entire time range. The  technique may, however, be
even more useful when doing predictions based on a model with time-dependent
covariates. A hazard model with time-dependent covariates $Z(t)$ for which the trajectories are still
unknown at $t=0$ can be useful when the aim of modeling is to describe the
processes behind the hazard,
but it is no longer simple to calculate the survival probabilities from the hazard using the relationship $S(t) = \exp(-\Lambda(t))$.
 This means that such a model cannot be used for predictions
at $t=0$. However, they can still be useful because the history of
$Z(t)$ before $t_{pred}$ can be informative for the future of the
process. Therefore, such predictions can
be based on landmark models. The history of $Z(t)$ up to
$t_{pred}$ is summarized in a single statistic that is used as
time-fixed covariate in the prediction model. The simplest approach is
to use the last observation before $t_{pred}$.
While this approach does not satisfy the consistency condition that a
prediction model at two different times should be compatible
\citep{jewell-nielsen, suresh-etal}, it can be extended to give better
predictions if more
flexible prediction models from the landmark time are used, and more
than just the last observation of $Z(t_{pred})$ is used to represent
the effect of $Z(t)$, including
e.g., cumulative effects \citep{keogh-etal}.

Another way is to develop a joint model for  $Z(t)$ and
$\al(t|Z(t))$ and estimate survival probabilities by conditioning
on  the history of $Z(t)$ at $t_{pred}$ and $T^* \ge t_{pred}$ in the
joint model. Estimation for such models can be challenging \citep{rizo-book}
and while such an approach has a better theoretical justification and
is efficient, there can be concerns about the robustness.

\subsection{Prediction with competing risks}
\label{sec:comprisk}
Very often intensity models are used for an event that does not
include all-cause mortality. This was for example the case in the lithium and
dementia study.
However, in the presence of competing risks, naively inserting the estimated hazard into equation
(\ref{eq:absrisk}) will produce an upwards biased estimate of the
absolute risk (cumulative incidence). This is because, by treating competing events as
censorings, one pretends that the target population is one where the
competing events are not operating and therefore neglects the fact
that subjects who have died from competing causes can no longer
experience the event of interest. In such a situation it is necessary also
to estimate the intensity of the competing events and to combine such
estimates with those for the event of interest into an estimate of the
\emph{competing risks cumulative incidence}. If the cumulative hazard for the
competing events is  $\Lambda_{02}(t|Z)$
then the cumulative incidence for a 1-event is given by
\begin{equation}
F_1(t\mid Z)=\int_0^t\exp(-\Lambda_{02}(u\mid Z)-\Al_{01}(u\mid Z))d\Al_{01}(u\mid Z).
\label{eq:cuminc}
\end{equation}
  Without covariates, the estimator obtained by plugging-in Nelson-Aalen estimates for the
  cumulative hazards in (\ref{eq:cuminc}) is known as the \emph{Aalen-Johansen estimator} \citep{ABGK-book}.\\

It is also possible to set up direct regression models for $F(t\mid
z)$, e.g., using the Fine and Gray regression model \citep{Fine99} but a further
discussion of such methods is beyond the scope of the present
paper. We will, however, exemplify the use of the cumulative incidence
in the examples in Section \ref{sec:examples}.

\section{Issues in causal inference}
\label{sec:causal}
`Causality' may be defined in a number of different ways  but the most commonly used
approach is based on potential outcomes and randomized experiments
\citep{rubin-book, goetghebeur, hernan-robin-book2020}.
This is because a well conducted randomized experiment allows a causal
interpretation of the estimated treatment effect. However, also in certain
observational studies a causal interpretation of the effect of a
non-randomized exposure is of interest. Two classic examples where
randomization cannot be employed are ($i$) post-marketing studies of
potential “adverse effects” of medications/treatments already approved
(based on earlier randomized trials that focused on their
effectiveness), and ($ii$) environmental or occupational exposures
(randomization often impossible and/or un-ethical). However, any
attempt of causal interpretation in an observational study, obviously,
requires strong assumptions. It is not the intention
of this paper to go into details concerning causal inference but
in the current section we will briefly discuss the topic. \\

First of all, causal questions are most natural and relevant for \emph{modifiable
  variables} for which a hypothetical randomized study could, in principle, be
done. They are less relevant for variables that you cannot change
(such as sex or race). Causal parameters are typically defined as contrasts
between average outcomes for the same population under the hypothetical
scenarios of every one being `treated' versus every one being
`untreated'. Thus, the causal risk difference at time $t$ is:
\begin{equation}
\Delta(t)=P(T^*(0)\leq t)-P(T^*(1)\leq t)
\label{eq:Delta}
\end{equation}
where $T^*(0), T^*(1)$ are the (possibly counterfactual) survival times
`under no treatment', vs. `under treatment'. The causal parameter,
$\Delta(t)$ in equation (\ref{eq:Delta}) is directly estimable in a randomized study and may be
estimable based on observational data under a set of assumptions,
including `no unmeasured confounders' -- a condition that can,
obviously, never be tested based on the available data.

Under these assumptions, one way of getting from a hazard model to an estimate of the counterfactual risk, had all subjects in the population
been treated with treatment $a=0,1$, is to use \emph{inverse probability of treatment weights}; another is to use the `$g$-formula'.
The latter works, as follows. If the hazard model for given treatment $A$ and confounders $Z$ leads to an estimated absolute risk of $\widehat{F}(t\mid A,Z)$ then the estimate of $P(T^*(a)\leq t)$ using the $g$-formula is
\begin{equation}
\widehat{P}(T^*(a)\leq t)=\frac{1}{n}\sum_i \widehat{F}(t\mid a,Z_i),\quad a=0,1.
\label{eq:g-formula}
\end{equation}
That is, the risk is predicted for each given subject under treatment
$A=a$ given his or her observed covariates and then \emph{averaged}
over the sample $i=1,\dots,n$. Formula (\ref{eq:g-formula}) is applied
separately for each treatment ($a=0$ vs $a=1$) and the estimate of
$\Delta(t)$ is obtained.
Note that the $g$-formula is useful for predicting average risk in a
sample even though a causal interpretation is not aimed at. We will
illustrate this is the examples of Section \ref{sec:examples}.

Another use of the $g$-formula can yield an estimate of the number of events `attributable to' a certain modifiable
risk factor, $A$. This number is given by the difference between the `total risk' \emph{observed} in the
population before time $t$: $\sum_i\widehat{F}(t\mid A_i, Z_i)$, and that expected \emph{if every one was unexposed} ($A_i=0$):
$\sum_i\widehat{F}(t\mid 0, Z_i)$. When doing this for a number of risk factors, these may be ranked in a way that also accounts for their prevalence in the population.\\

 Though a hazard model may, thus, be useful both for describing associations between covariates and a time-to-event outcome and serving as a useful step towards estimating a causal contrast like (\ref{eq:Delta}), it does not itself provide a causal contrast. This may be seen, as follows.
Recall the intuitive definition (\ref{eq:int}) of the hazard function for survival
data:

$$\lambda(t)=P(T^*\leq t+dt\mid T^*>t)/dt.$$
This shows that contrasts based on the hazard functions for the counterfactual outcomes $T^*(0),T^*(1)$, e.g. the hazard ratio at time $t$, $$\frac{\lambda^1(t)}{\lambda^0(t)}=\frac{P(T^*(1)\leq t+dt\mid T^*(1)>t)}{P(T^*(0)\leq t+dt\mid T^*(0)>t)}$$
are not directly causally interpretable since they contrast different
sub-populations: those who survive past $t$ under treatment ($T^*(1)>t$)
and those who survive past time $t$ under no treatment ($T^*(0)>t$). For
this reason, a statement saying that `treatment only works until time
$\tau$ but not beyond' in a situation with $\beta(t)<0 \mbox{ for }t<\tau$ and
$\beta(t)=0 \mbox{ for }t>\tau$ is not justified \citep{hernan-EPIDEMIOLOGY2010,
            aalen-LiDA2015, torben-etal}.

A special problem with causal inference for survival data is
\emph{time-depending confounding/mediation} where a time-dependent covariate both
affects future treatment and survival outcome and is affected by past
treatment. For this situation, special techniques are needed to draw
causal conclusions concerning the treatment effect \citep{daniel-etal,
  hernan-robin-book2020}.

\section{Illustrative applications}
\label{sec:examples}

In this section, we illustrate the points given in the paper with three real data examples. A more detailed analysis (along with code in R statistical software) is provided in the online Appendix.

\subsection{Peripheral arterial disease}

Peripheral arterial disease (PAD) is a common circulatory problem in which
narrowed arteries reduce blood flow to peripheral limbs, often the legs.
It is also likely to be a sign of a more widespread atherosclerosis, and
subjects manifesting the disease carry an increased risk
for atherothrombotic events.
The PAD data set contains the results of a Slovene study reported in \cite{Blinc17, Blinc11}.
Briefly, the study was conducted by 74 primary care physicians-researchers (GPs),
who recruited subjects with PAD along with age and
sex matched `controls' without PAD. Yearly examination visits were planned with a total of 5 years of follow-up.
The final study included 742 PAD patients and 713 controls, with
baseline data for each subject, measurements at each visit, and endpoints.
Important endpoints are death, either due to cardiovascular disease (CVD)
or other causes, non-fatal CVD endpoints of infarction and stroke, and patient
interventions attributed to the disease such as re-vascularization procedures.
All the individuals in the study were treated according to the latest treatment guidelines and the goal of the study was to study survival of the patients with PAD (in comparison to controls) despite optimal treatment.\\

\noindent \textit {Endpoints:} Most of this analysis will focus on death as the outcome of interest. The causes of death are split into two groups: cardiovascular (CV) or other (non CV) and, in addition,  we will also consider all CV events (stroke, infarction or death) as an outcome for modelling. \\

\noindent \emph{Inclusion criteria, follow-up and censoring:} With most GPs, the follow-up visits of the patients followed a yearly plan, though, in practice, the visits tended to be moderately delayed, with time to the 5th visit ranging from 4.8 to 6.8 years. Most data on patients were recorded at the time of their visit, with the exception of deaths which were reported as they occurred (along with all other events that occurred since the last visit).

Because of between-physician differences in whether patients were
followed after 5 years, to avoid possibly non-independent censoring, all
individuals alive at 5 years after enrollment were censored at that
time. \\

\begin{figure}\centering
	\includegraphics[height=5cm]{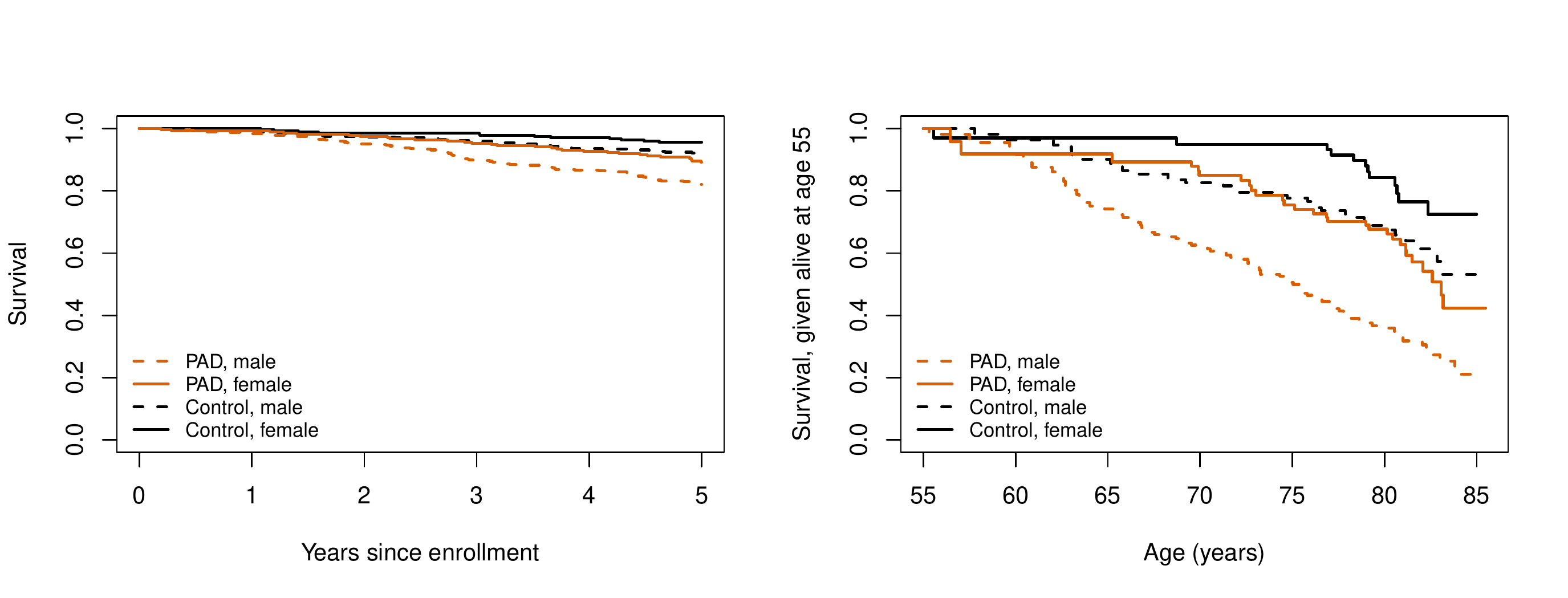}
	\caption{Kaplan-Meier estimates of cumulative probability of survival with respect to time since enrollment (left) and age (right).}
	\label{fig:pad1_km}
\end{figure}

\noindent \emph{Time axis, basic survival analysis:}
For the PAD patients the time since diagnosis is a natural time axis as
it represents both the progression of disease and treatments for the
disease.  Survival curves for the control subjects serve as a comparison
outcome of similarly aged subjects without the disease, but do not have
a natural stand-alone interpretation. Figure \ref{fig:pad1_km} contains the overall Kaplan-Meier curves for PAD and control subjects, male and female (deaths of any cause are considered as the outcome).
The survival is higher for females than for males, which is no surprise
given a mean age at entry of 65 years, and is lower for PAD subjects than
for the age and sex matched controls.
The right hand panel shows the curves on age as the time axis, a very similar pattern can be observed. When using age, the left hand portion of a survival curve can often be highly variable due to the small number of the patients at risk at a young age and this early high variability can then affect the entire curve. To avoid this, we estimate conditional survival $P(T^*>t | T^*>t_0)$ for $t>t_0$, with $t_0$ chosen so that the risk set is large enough and most of the information of interest is included. In the case of PAD, we choose $t_0=55$ years.\\

\noindent\emph{Hazard regression models}

\noindent We next study how the covariates affect the hazard of dying.

\begin{figure}\centering
	\includegraphics[height=5cm]{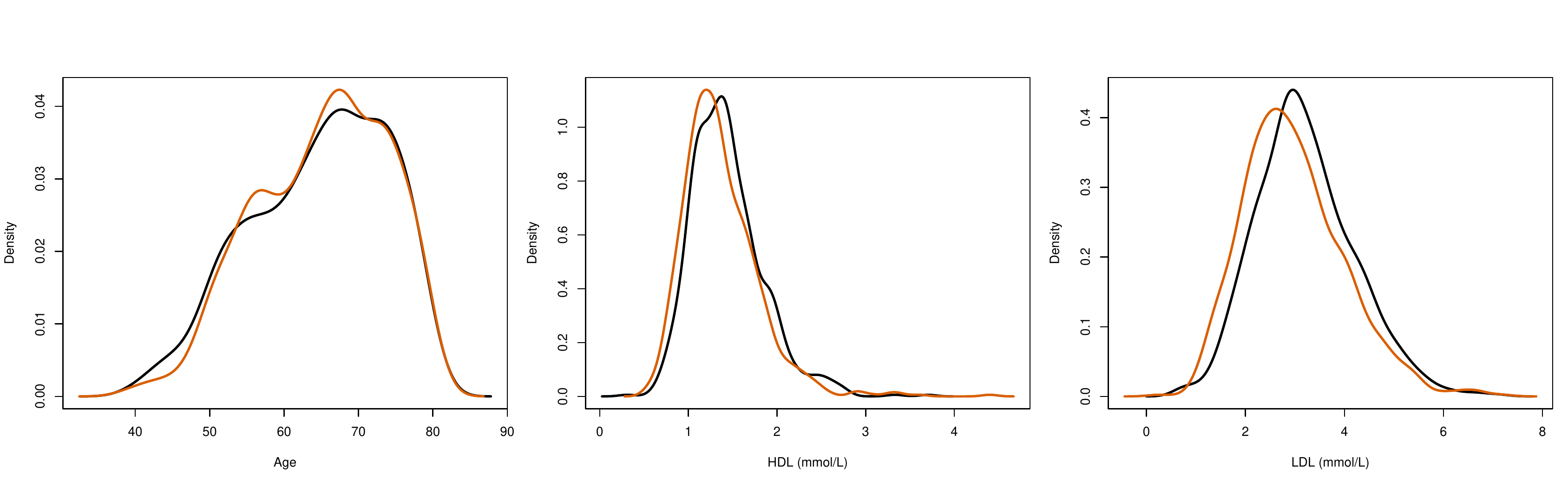}
	\caption{Distributions of continuous variables with respect to PAD (red=PAD, black=Control).}
	\label{fig:pad2_cov}
\end{figure}

\noindent\emph{Covariates:} We will be interested in PAD, sex (38\% women), age, and later also in LDL and HDL. The distribution of the continuous covariates with respect to PAD is given in Figure \ref{fig:pad2_cov}. By study design there should be no difference in the age distribution, HDL and LDL are slightly lower for the PAD subjects. When used as a covariate, age will be expressed in decades to give a coefficient of a more interpretable size.

In the analysis presented in Table \ref{tab:padfit}, we focus on the effect of PAD and the effect of sex in each PAD subgroup.
First, we fit a Cox model with time since enrollment as the time axis. Knowing that age is a strong predictor, we include it in the model (i.e., age/10 is used as a covariate). We learn that patients with PAD have a 2.4 times higher hazard than the controls,
and that male sex and 10 additional years of age each increase the hazard by approximately 2 fold. The effect of both sex and age is very similar in both groups (patients and controls).\\
 Alternatively we can use age as the time axis and add time since enrollment (FU time) as a possible predictor. Using age as the time axis the effect of male sex is a 2-fold hazard increase,
just as before.
The time-dependent variable years-since-enrollment compares the hazard of death for
subjects with more study years to those recently enrolled, and shows an increase
in death rates over time (HR=1.2 per year).

By choosing one time axis and adding the other into the model as a
covariate, the interpretation of the HR for sex becomes equal (the HR
for two patients of the same age and same time since enrollment) under
the condition that the assumptions of the linearity (and PH) of the
covariate are met.
To avoid the problem of choosing the time axis and adding assumptions,
the Poisson model that allows multiple time axes can be used
instead. To this end, we assume the baseline hazard constant within
yearly intervals of time since enrollment and five-year intervals of
age. The results are given in the last two rows of Tablec
\ref{tab:padfit}. We can see that, in our case, all three approaches
coincide well, so, the possible violations of the assumptions of the
different options had no effect.\\

\begin{table} \centering
  \begin{tabular}{lccccccc}
    & \multicolumn{2}{c}{Overall} & \multicolumn{2}{c}{PAD} &
    \multicolumn{2}{c}{Control} & p \\
    & HR & 95\% CI & HR &  95\% CI & HR & 95\% CI & PAD vs C\\
    \hline
  \multicolumn{8}{c}{Time since enrollment axis, Cox model}\\
PAD& 2.40 &(1.71, 3.37) \\
Sex (m vs. f)& 2.00 &(1.40, 2.86) &2.01 &(1.31, 3.08) &1.97 &(1.02, 3.79) &0.96 \\
Age (per10yrs)& 1.93 &(1.57, 2.37) &1.91 &(1.49, 2.45) &1.98 &(1.36, 2.89) &0.88 \\
\\ \multicolumn{8}{c}{Age axis, Cox model}\\
PAD& 2.40&(1.70, 3.37) \\
Sex (m vs. f)& 2.02 &(1.42, 2.90) &2.01 &(1.31, 3.08) &2.01 &(1.04, 3.88) &1.00 \\
FU (per1yr)& 1.18 &(1.05, 1.33) &1.20 &(1.05, 1.38) &1.12 &(0.91, 1.39) &0.61 \\
\\ \multicolumn{8}{c}{Both time axes, Poisson model}\\
PAD& 2.38&(1.70, 3.35) \\
Sex (m vs. f)& 2.01 &(1.41, 2.88) &2.03 &(1.33, 3.11) &1.97 &(1.02, 3.81) &0.95 \\
\end{tabular}
\caption{Estimated hazard ratios (HR) and 95\% confidence intervals (CI) in models with different time axes, fitted with Cox or Poisson model. The last column reports the p-value for interaction of each covariate with group (PAD or control).}
\label{tab:padfit}
\end{table}

\begin{figure}\centering
	\includegraphics[height=5cm]{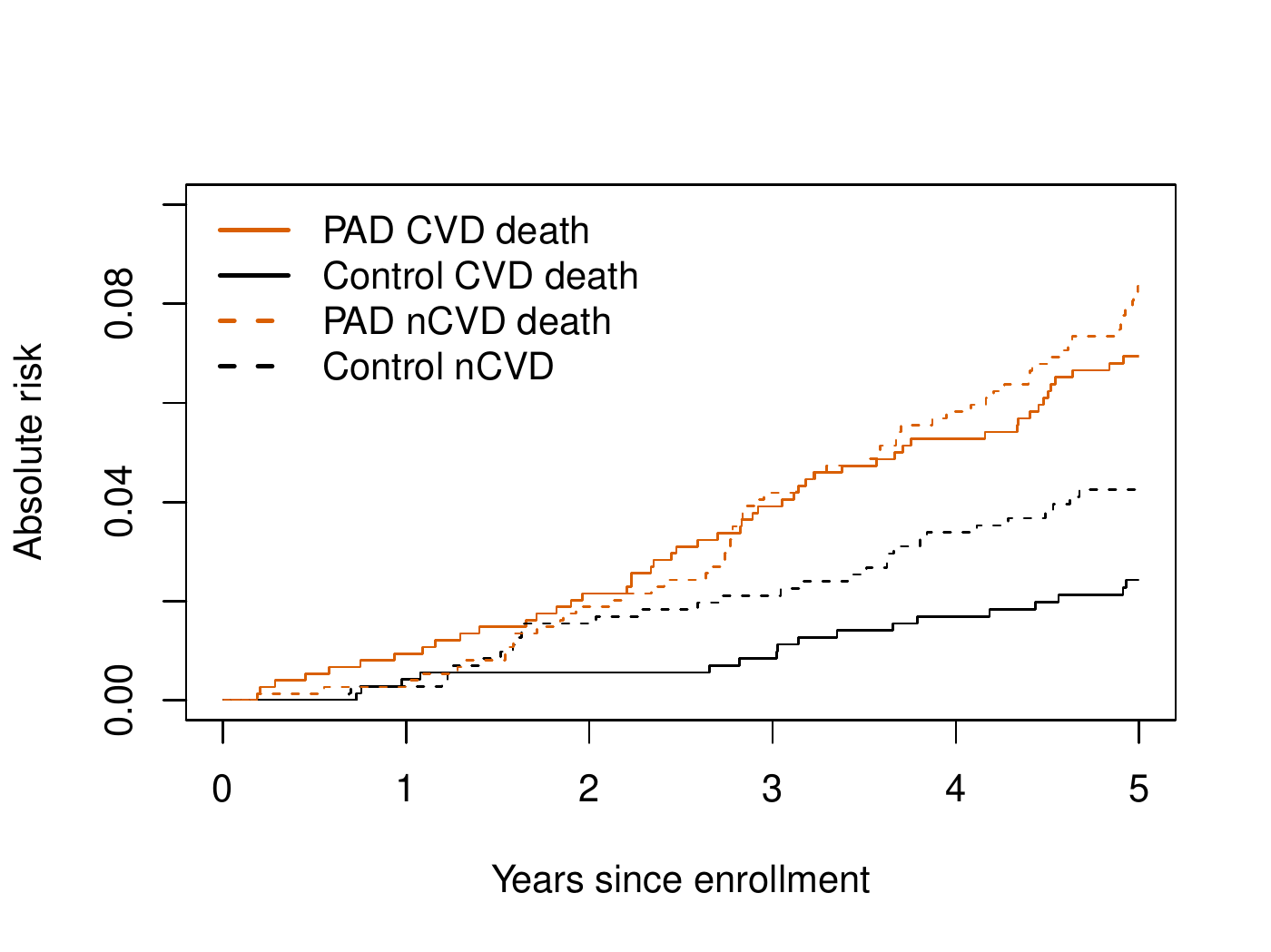}
	\caption{The probability of dying due to cardivascular (solid line) or other reasons (dashed line) with respect to PAD (red=patients, black=controls) }
	\label{fig:pad3_AJ}
\end{figure}

\noindent\emph{Competing risks and time-dependent covariates}\\
The analysis so far considered all causes of death equally, but the
cardiovascular deaths (CV) are of particular interest. In 5 years, 159
patients died, 68 of these due to cardiovascular reasons. Figure
\ref{fig:pad3_AJ} presents the Aalen-Johansen estimator of the
absolute risk (also known as the cumulative incidence function). The
estimated 5-year survival probability of PAD patients is 84.5 \% and
we can see that 6.9\% of the PAD patients are estimated to have died
due to CV reasons and 8.4 \% due to other reasons. Both the
probability of CV but also that of non-CV death are considerably
greater than in the control group.\\

We wish to explore the effect of our basic covariates (sex, age and
PAD) and cholesterol (LDL, HDL) on the hazard for cardiovascular death
or major cardiovascular events.
Hazard models for a particular endpoint can be fitted by censoring all
`other cause' deaths.  The results for Cox models on the
time-since-enrollment axis are given in Table \ref{tab:cmp_tdc}.
\begin{itemize}
\item A: As before, we see that male sex and higher age increase the
  hazard, we also see that PAD is a strong risk factor, the hazard of
  PAD patients is almost 3 times higher than that of the
  controls. Neither LDL nor HDL values at baseline seem to have an
  important effect.
\item B: We now include all the information available - the values of HDL and LDL were updated on a yearly basis (if missing, the last value was carried forward), i.e., we use them as time-dependent variables. We can now observe a much larger effect of HDL,  patients whose HDL is lower by 1 mmol/l have a 4.7 ($\approx 1/0.21$) times higher hazard. We can also observe a more pronounced effect of LDL. Its direction may seem counterintuitive, but may be due to the fact that patients at a higher risk have lower target values of LDL and hence the lower LDL may be a proxy for the higher risk patients.
\item C: This model regards not only CV death but also stroke and infarction as events.  The effects of the covariates do not change much, but all the standard errors have decreased as the number of events increased to 142.
\end{itemize}

\begin{table} \centering
  \begin{tabular}{l|cc|cc|cc|}
  & \multicolumn{2}{|c|}{A}& \multicolumn{2}{|c|}{B} & \multicolumn{2}{|c|}{C} \\
& \multicolumn{2}{|c|}{CV death} & \multicolumn{2}{|c|}{CV death} &\multicolumn{2}{|c|}{CV events}\\
& \multicolumn{2}{|c|}{Time-fixed}& \multicolumn{2}{|c|}{Time-dependent} & \multicolumn{2}{|c|}{Time-dependent}\\ \hline
& HR &95\%CI& HR&95\%CI& HR &95\%CI \\
\hline
PAD&2.87&(1.65-5)&2.40&(1.37-4.20)&2.27&(1.57-3.28)\\
Sex (m vs. f)&1.67&(0.97-2.88)&1.36&(0.79-2.35)&1.90&(1.28-2.81)\\
Age (per10yrs)&1.93&(1.40-2.66)&2.01&(1.45-2.77)&1.54&(1.25-1.90)\\
HDL (mmol/l)&0.74&(0.39-1.41)&0.21&(0.10-0.48)&0.48&(0.29-0.79)\\
LDL (mmol/l)&0.92&(0.72-1.18)&0.76&(0.57-1.01)&0.88&(0.73-1.07)\\
\end{tabular}
\caption{Estimated hazard ratios (HR) and 95 \% confidence intervals (CI) in Cox models. A: CV deaths, baseline HDL and LDL; B: CV deaths, time-dependent HDL and LDL; C: all CV events, time-dependent HDL and LDL.}
\label{tab:cmp_tdc}
\end{table}

\begin{figure}\centering
	\includegraphics[height=5cm]{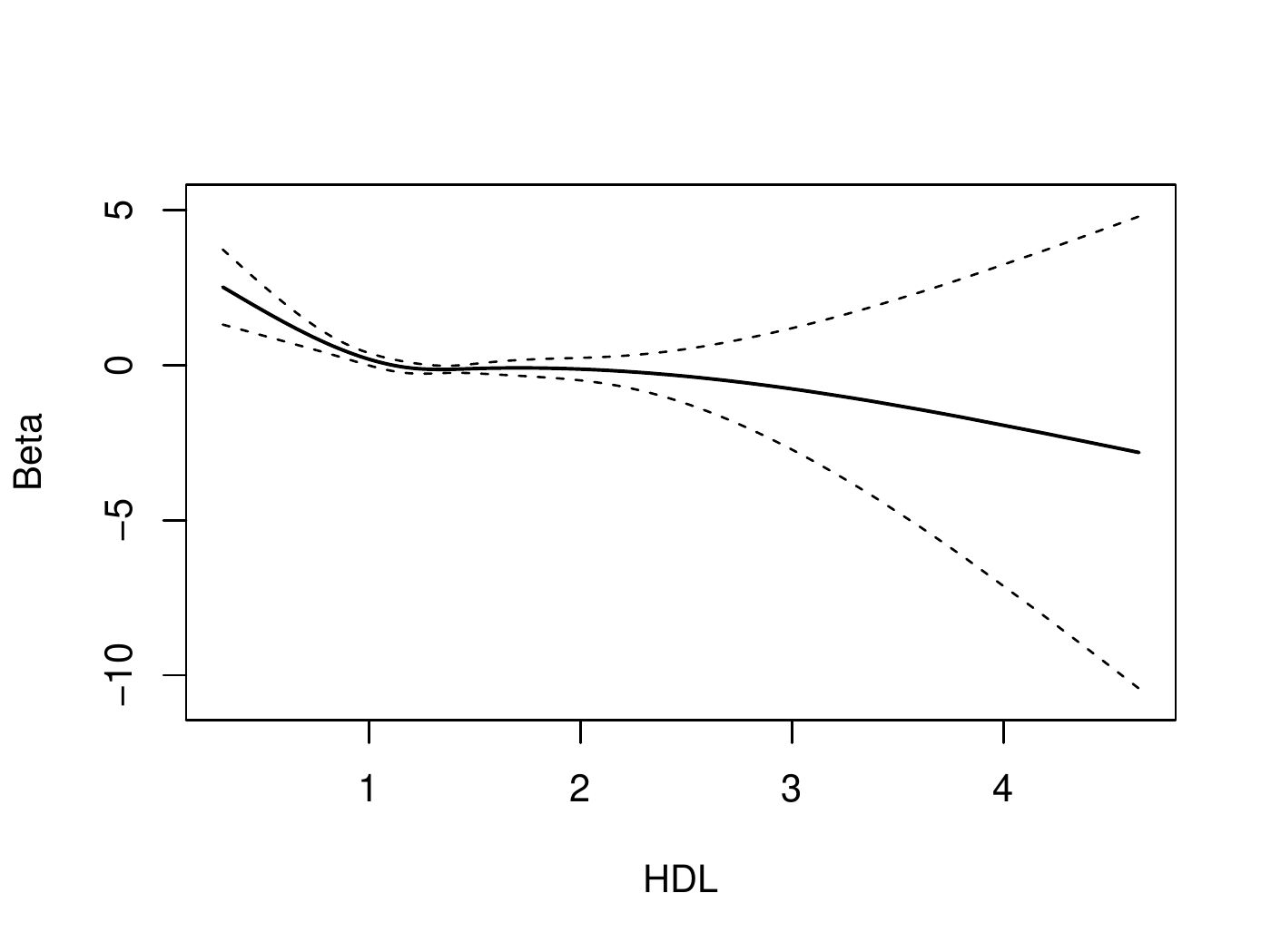}
	\caption{The effect of HDL on the hazard of having a CV event modelled by using restricted cubic splines in a multiple Cox regression model (an extension of model C). }
	\label{fig:pad4_nlin}
\end{figure}

To check whether the above interpretation makes sense, we further
examine the goodness-of-fit of the models. \\
We focus on model C, which uses all the information available. Adding
a spline to the model, i.e. replacing $\beta$ HDL  by s(HDL), we can
see that the linearity of HDL may be problematic. The protective
effect increases with the value of HDL but may level off for values
above approx 1.5 mmol/l, see Figure \ref{fig:pad4_nlin} (the huge
confidence interval beyond 2 mmol/l is due to very few individuals
with HDL above 2). Allowing HDL to be non-linear, the Schoenfeld's
residuals test for proportional hazards indicates no further issues.

\begin{figure} \centering
	\includegraphics[height=5cm]{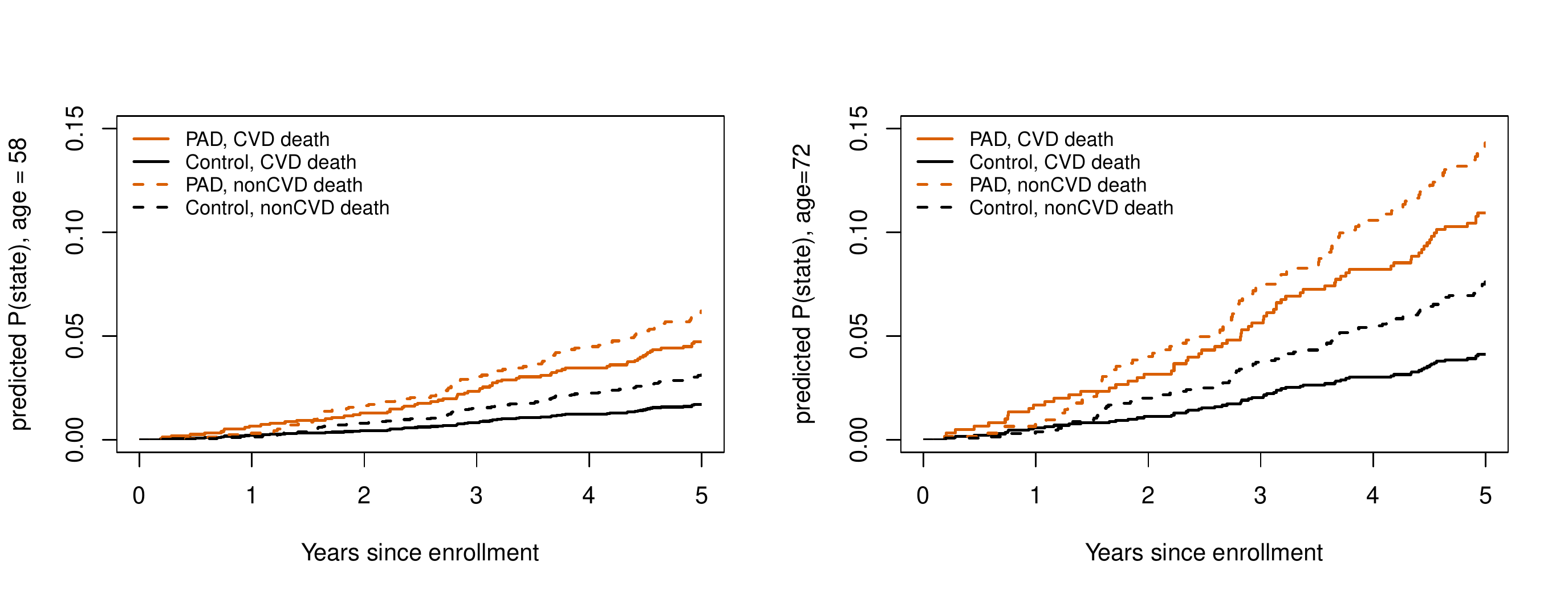}
	\caption{The absolute risk of dying of CV (solid line) and other causes (dashed line) by PAD status (patients=red, controls=black) and age (left graph 58, right graph 72). The values of baseline HDL and LDL are 1.3 and 3, respectively.}
	\label{fig:pad5_AR}
\end{figure}

The calculation of the absolute risk is based on model A, as only the
baseline information can be used for prediction. Unlike in the case of pure
hazard modeling, the other causes of death cannot be simply censored - to estimate
the probability of dying due to cancer, the hazard of dying due to
other causes must be estimated as well, see Table
\ref{tab:cmp_tdc2}. The absolute risks of two individuals, one aged 58
and the other 72 (25th and 75th percentile of age) and median values
of lipids are plotted in Figure \ref{fig:pad5_AR}.

\begin{table} \centering
  \begin{tabular}{r|cc}
    &\multicolumn{2}{c}{Time-fixed, other cause} \\
    & HR &95\% CI  \\ \hline
    PAD & 2.04 & (1.31--3.19) \\
Sex (m vs. f) & 2.12 & (1.29--3.50) \\
Age (per 10 yrs) & 1.93 & (1.45--2.56) \\
HDL & 0.82 & (0.43--1.55) \\
LDL & 1.02 & (0.83--1.26) \\
  \end{tabular}
  \caption{Estimated hazard ratios (HR) and 95\% confidence intervals (CI) for other cause mortality. Baseline HDL and LDL.}
  \label{tab:cmp_tdc2}
\end{table}

To conclude, we have seen that the PAD remains a strong risk factor
despite following the latest treatment guidelines. This is true
regardless of whether we focus on all major cardiovascular events or
on CV death only. Old age and too low HDL are further associated with
a higher event rate.

\subsection{Non-alcoholic fatty liver disease}

Non-alcoholic fatty liver disease (NAFLD) is defined by three criteria:
presence of greater than 5\% fat in the liver (steatosis),
absence of other indications for the steatosis (such as excessive
alcohol consumption or certain medications), and absence of other liver
disease \citep{Puri12}.
   NAFLD is currently believed to be responsible for almost 1/3 of
liver transplants and its impact is growing. It is expected to be a major
driver of hepatology practice in the coming decades \citep{Tapper18}.
The study included all patients with a NAFLD diagnosis in Olmsted County,
Minnesota between 1997 and 2014 along with up to four age and sex matched
controls for each case \citep{Allen18}.
(Note that some changes to the public data have been made to protect patient
 confidentiality; analysis results here will not exactly match the original paper).

The goal of the study is to investigate
whether NAFLD subjects are at increased mortality risk compared to the general
population, and if so the amount of increase. Only a minority of subjects are tested for NAFLD since this requires an abdomnial scan, and we can, therefore, only address the progression of \emph{detected} NAFLD.\\

\noindent \emph{Entry time, inclusion criteria: }
In the PAD study, the data were collected prospectively and hence the
inclusion criteria were naturally evaluated at the time of
inclusion. On the contrary, the NAFLD data were collected
retrospectively using existing databases and are hence more prone to
mistakes regarding the time when the inclusion criteria are known.

Subjects enter the study at the age of NAFLD diagnosis or selection as a
control, whichever comes first.
Because NAFLD is often a disease of exclusion,
a NAFLD diagnosis followed shortly by the diagnosis of another liver
disease is considered a false positive.  The data set is restricted to
`confirmed NAFLD', i.e., if someone were diagnosed on 2001-06-20, the index date for confirmed NAFLD would be 2002-06-20, assuming that another liver diagnosis, death, or incomplete follow-up did not intervene.
The follow-up of the matched control subjects also commences on the `confirmed NAFLD' date.
This is important.  If the matched subjects' follow-up were started on 2001-06-20
then the control has the opportunity to die during that first year while the
case does not, leading to immortal time bias.

When selecting the controls for any given NAFLD case at age $a$, it is very
important only to use information that was available at age $a$ for the
controls.  We cannot exclude subjects who have too short a follow-up (die or
censored before age $a+2$ say), will later have diabetes, or, most particularly,
those who will later become NAFLD patients.
Each of these is a variant of immortal time bias.
In this data set, 331 of the subjects selected as controls were diagnosed with
NAFLD at a later age. Care must be taken at the time of analysis to correctly deal with these patients.
The preliminary checks and figures will treat each subject's value at study entry as fixed,
the hazard models will treat it as a time-dependent covariate.\\

\noindent \emph{Endpoints and censoring:} The primary focus of this analysis is death,
which means the endpoint is not problematic.
All the subjects in the study are administratively censored at the end of 2017, when the data set was created.
A small number  has been censored due to migration, about 1\% per year over the
age of 50 \citep{Sauver12}.

Because the publicly available NAFLD data set does not contain dates,
a plot of the censoring distribution is not particularly informative:
we do not know what the result \emph{should} look like.

Since the follow-up is as long as 20 years for some subjects, care
must be taken with the independent censoring assumption - the later
included subjects have a systematically lower death rate, e.g. due to
improved
general population medical care, and are, due to the later inclusion
date, followed-up for a shorter time period. \\

\noindent \emph{Time axis, basic survival analysis:}
The NAFLD is a not an acute condition, it may well exist for many
years before detection. Furthermore, the age range in the study is
very wide, death as the primary endpoint is highly related to age and
cases and controls match with each other on age (with 'time since
NAFLD' not well defined for the controls). All these reasons make age
the natural time axis for the NAFLD study.  This approach mimics an
idealized (but impractical) study which
included the entire population from birth forward, with time dependent
NAFLD as the covariate.

\begin{figure}\centering
	\includegraphics[height=5cm]{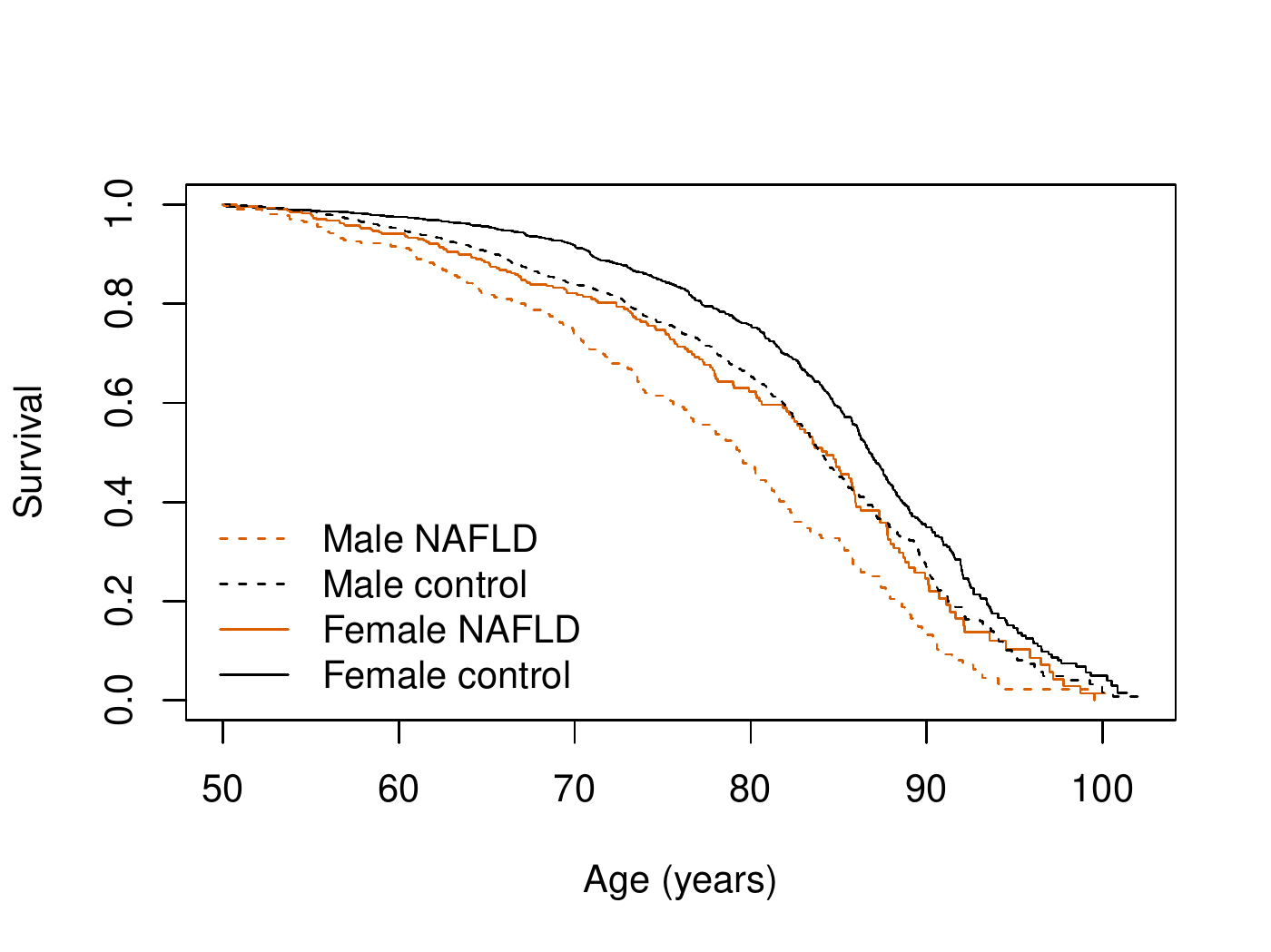}
	\caption{Survival curves from age 50 forward, comparing NAFLD to
non-NAFLD at study entry, stratified by male/female.}
	\label{fig:nafld1_surv}
\end{figure}

As a first description of data, we plot the estimated survival curves
for the patients by sex and NAFLD group, see Figure \ref{fig:nafld1_surv}.  For the latter we use
the subject's NAFLD status at enrollment as a time-fixed variable.
This approach is similar in spirit to using intent-to-treat in a clinical
trial, in that it gives a reliable estimate but one that may underestimate the
true clinical effect of a covariate. As with the PAD study we estimate
conditional survival.

\begin{figure} \centering
  \includegraphics[height=5cm]{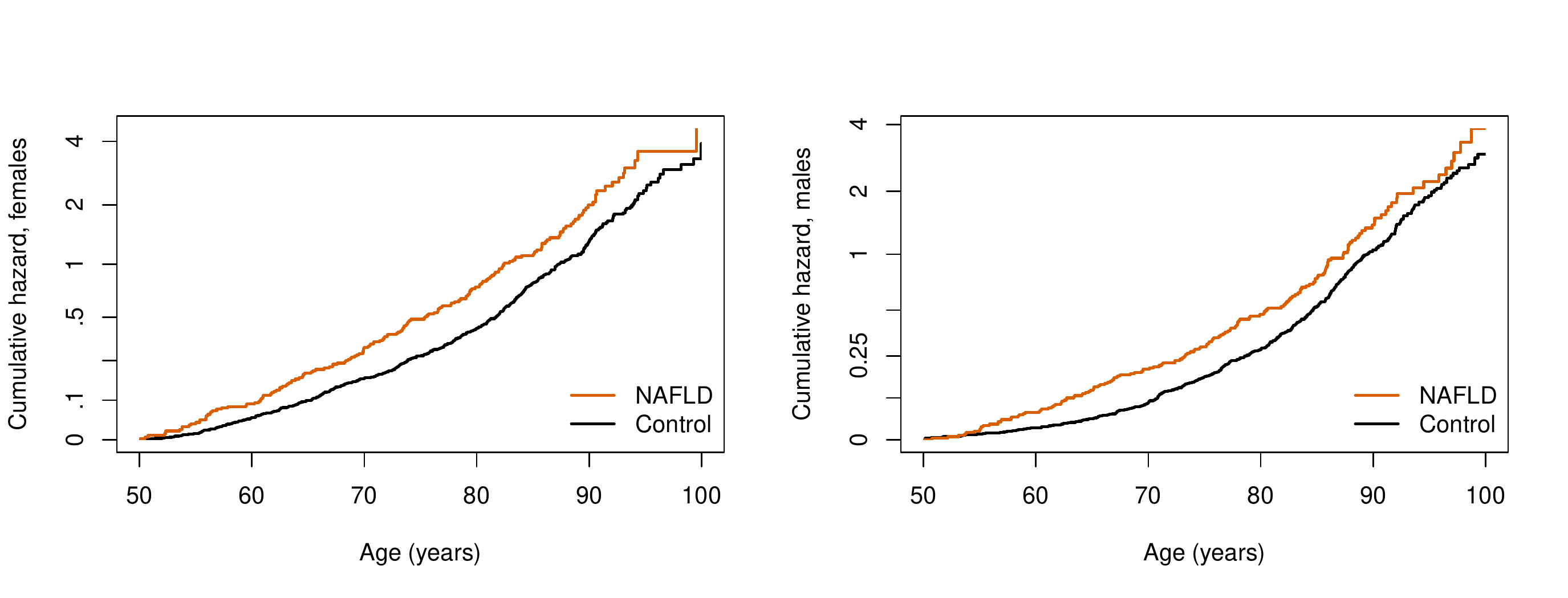}
  \caption{Nelson-Aalen estimates for the cumulative hazard from age 50, stratified by gender (left: females, right: males) and NAFLD (NB: log-scale on the vertical axis).
}
  \label{fig:nafld2_na}
\end{figure}

An alternative summary is to report the cumulative hazard (using the Nelson-Aalen estimator) by sex and time-dependent NAFLD status, see Figure \ref{fig:nafld2_na}. 
If the hazard is constant on the interval, the increase of the cumulative hazard on each interval is close to the death rates (proportion of deaths per person-year) given in Table \ref{tab:nafld_py}.
The hazard ratio (difference on log scale) between male patients and controls
is nearly constant in time, which suggests a proportional hazards model may fit
well.
 On the other hand, in women the NAFLD/control hazard ratio is highest at the
youngest ages and decreases with age. \\

\begin{table} \centering
  \begin{tabular}{l|c|c|c|c|}
&Female control&Female NAFLD&Male control&Male NAFLD \\ \hline
40-50 & 1.3&2.4&2.2&2.5 \\
50-60 & 2.5&5.9&5.2&8.3 \\
60-70 & 5.4&14.8&11.6&22.8 \\
70-80 & 18.0&28.1&23.4&37.2 \\
80-90 & 68.1&76.3&79.6&108.4 \\
\end{tabular}
\caption{Death rates per 1000 person years, split by age group, sex, and time-dependent NAFLD status.
}
\label{tab:nafld_py}
\end{table}

\noindent\textit{Hazard models:}\\
In the hazard models we can incorporate NAFLD as a time dependent covariate.
Subjects who were NAFLD at enrollment have a value of 1, controls start with
a value of 0 at enrollment, some of whom turn to 1 when they are diagnosed
with NAFLD at a later age.
Also, the overall model is not adversely affected by the small sample issue at
younger ages, so there is no need to use a restricted age range.
Because NAFLD is strongly associated with obesity, we also fit models that adjust
for other conditions associated with obesity: diabetes,
hypertension and dyslipidemia.
Fits were done overall and for males and females separately.

\begin{table} \centering
   \begin{tabular}{l|cc|cc|cc|}
 & \multicolumn{2}{|c|}{Overall}  & \multicolumn{2}{|c|}{Females} & \multicolumn{2}{|c|}{Males} \\
& HR &95\%CI& HR&95\%CI& HR&95\%CI\\ \hline
NAFLD only & 1.62 & (1.44--1.82) & 1.65 & (1.39--1.95) & 1.60 & (1.35--1.88) \\ \hline

NAFLD & 1.43 & (1.26--1.62) & 1.39 & (1.17--1.67) & 1.45 & (1.22--1.73) \\
Diabetes & 1.77 & (1.57--2.01) & 1.94 & (1.62--2.32) & 1.64 & (1.38--1.94) \\
Hypertension & 1.24 & (1.08--1.42) & 1.33 & (1.10--1.63) & 1.16 & (0.96--1.41) \\
Dyslipidemia & 0.68 & (0.60--0.78) & 0.65 & (0.54--0.79) & 0.72 & (0.60--0.88) \\
\end{tabular}
   \caption{Estimated hazard ratios (HR) and 95 \% confidence intervals (CI) from Cox models that have only NAFLD as a covariate,
     and models with NAFLD and covariates.  The overall model is fit to all subjects
     with sex as a stratification variable.
   \label{tab:nafld2}
}
\end{table}

Table \ref{tab:nafld2} contains the estimated hazard ratios from Cox models that include
all subjects, only males or only females, and for models that
include only NAFLD as a (time-dependent) covariate as well as adjusting for confounders.
The estimated effect of NAFLD is attenuated when adjusting for
the three covariates.
The higher prevalence of diabetes and other conditions explains a portion of the
NAFLD effect.  The overall NAFLD effect does not differ markedly for
males and females.\\

\begin{figure}\centering
  \includegraphics[height=8cm]{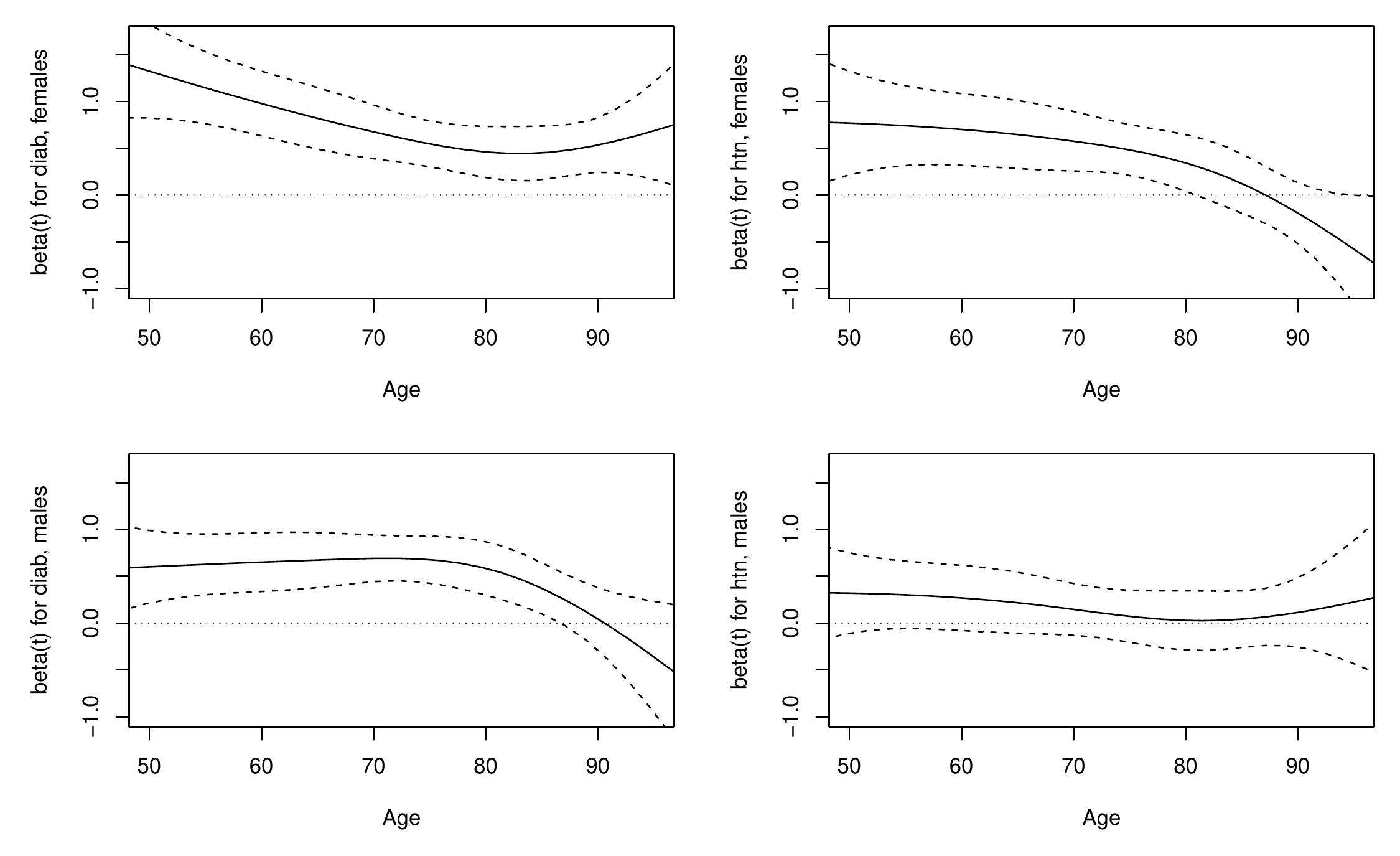}
\caption{The changing effect of diabetes (diab) and hypertension (htn) in the multiple regression models. Top row: females, bottom row: males}
\label{fig:nafld3_nph}
\end{figure}

\noindent\textit{Model checks:}\\
Since all of the covariates in the models are binary,
there is no need to explore functional form. An overall test for proportional hazards based on Schoenfeld's residuals has results that mimic what
was seen in the hazard plot of Figure \ref{fig:nafld2_na}: males fit the
proportional hazards model well (p=.4) while females have significant
non-proportionality ($p<0.001$).
It is interesting that the overall `average' effects, over age, are
nearly the same for male and females, however.
Checks of the multiple Cox model show that non-proportionality is more
severe with respect to diabetes (for both males and females) and for hypertension for women, see Figure \ref{fig:nafld3_nph}.
The relative effect of comorbidities on death rates is higher at younger ages, but may get high again with very old age.

In conclusion, NAFLD is associated with an increased mortality compared to disease-free,
age and sex matched controls. Part of this increase may be explained by the different diabetes, hypertension and dyslipidemia distributions in the two groups.

\subsection{Advanced ovarian cancer}
As the last example, we consider the advanced ovarian cancer data set.
It contains follow-up on 358 subjects who
were enrolled
in two trials of chemotherapy for advanced ovarian cancer, conducted
around 1980 by a multi-institution
research network in the Netherlands.
The eligibility criteria for enrollment included
pathologic confirmation of advanced disease, age less than 70 years,
lack of serious cardiac or renal disease, and favorable haematological status.
Patients could not have a second tumor, brain metastases, or prior
radiation or chemotherapy.
The treatment is not of our main interest here, hence we treat this data set primarily as an observational study.
The data were extensively analyzed in Chapter 6 of \cite{van12}, and
further references can be found there as well. Patient follow-up in the data set continued for 6 years. The goal of the analysis was to predict the survival probability of
patients using covariates that were recorded at baseline. \\

\begin{figure}\centering
  \includegraphics[height=5cm]{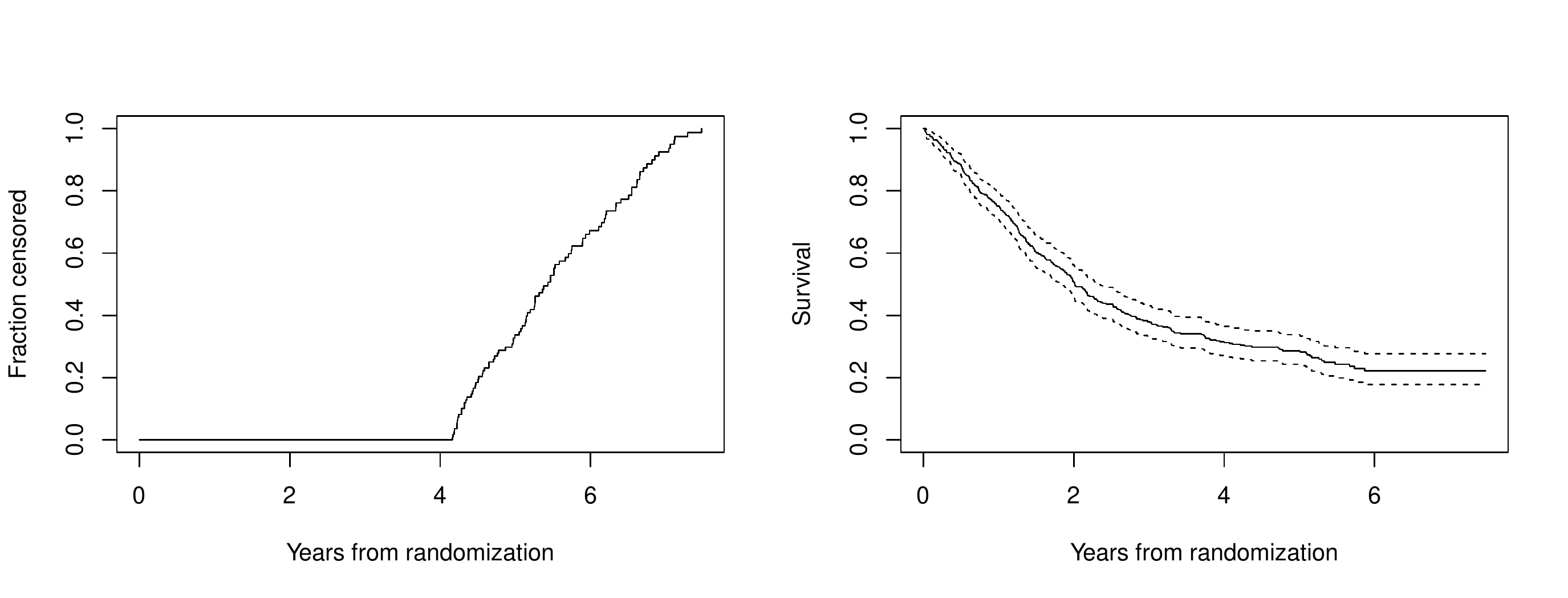}
\caption{Censoring fraction and survival curve (with 95\% confidence interval) for the ovarian cancer study.}
\label{fig:ovarian1_censurv}
\end{figure}

Focusing on a fatal condition such as advanced cancer in a data set
that comes from a clinical trial with excellent follow-up, the basic aspects of
these data are particularly simple: the sole event of interest is death
of any cause, the inclusion criteria are clear and the most natural
time axis is time from diagnosis as this is the time frame of most
direct interest to both the patient and the care provider. The left
panel of figure \ref{fig:ovarian1_censurv} shows the censoring pattern
for the study, which follows the expected `hockey stick' shape for a
formal trial with 3 years of enrollment, 4 years of follow-up after
enrollment of the final subject, and no subjects lost to
follow-up. The graph shows no censoring before 4 years followed by an
upward line corresponding to uniform accrual each year. The
Kaplan-Meier curves give the overall pattern of survival for this cohort,
see the right hand graph of Figure \ref{fig:ovarian1_censurv}. \\

\noindent\emph{Regression:}
In our analysis, we focus on three covariates:

\begin{itemize}
  \item FIGO: This is a staging system for ovarian cancer.
    Advanced ovarian cancer comprises the stages  III ($n=262$, used as reference group) and IV ($n=96$).
    Stage IV patients are known to have a very poor prognosis.
   \item The diameter of the residual tumor after surgery, categorized as
     micro, $< 1$ cm, 1--2 cm, 2--5 cm, and $>5$ cm, with the last category being the most frequent one ($n=145$). We will use the 'micro' category ($n=29$) as the reference group.
  \item Karnofsky index. A measure of the patient's functional status at
    the time of diagnosis.  The maximum score 10 is an indication of no
    physical limitations. We will regard the covariate as quantitative in the model.
\end{itemize}

The coefficients in the fitted PH model are given in Table \ref{tab:ovarian1}.\\

\begin{table}\centering
   \begin{tabular}{l|cc}
 & HR & 95\% CI \\ \hline
Diameter$<$1cm & 1.38 &(0.73-2.57) \\
Diameter 1-2cm & 2.24 &(1.19-4.21) \\
Diameter 2-5cm & 2.38 &(1.29-4.39) \\
Diameter$>$5cm & 2.53 &(1.40-4.57) \\
FIGO (stage IV vs. III) & 1.73 &(1.33-2.25) \\
Karnofsky index (per 1 point)& 0.84 &(0.75-0.93) \\
\end{tabular}
   \caption{Estimated hazard ratios (HR) and 95 \% confidence intervals (CI) in the Cox model for the ovarian cancer data. The reference group for the covariate Diameter is `micro'.
   \label{tab:ovarian1}
}
\end{table}

\begin{figure}\centering
  \includegraphics[height=5cm]{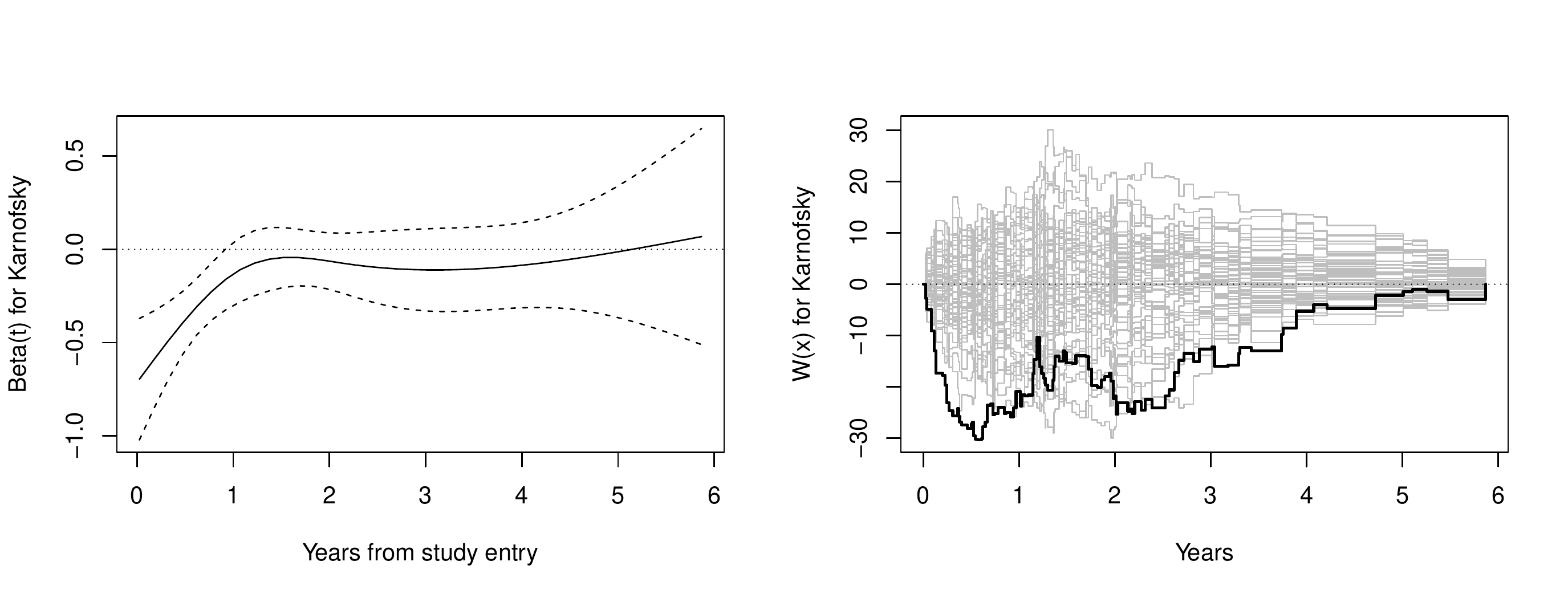}
\caption{Proportional hazards plots for Karnofsky score (left: smoothed Schoenfeld residuals, right: cumulative Lin et al )}
\label{fig:ovarian2_PH}
\end{figure}

\noindent\emph{Checking the assumptions of the model:}\\
To check the proportional hazards assumptions, we consider both the
method using cumulative Schoenfeld residuals of \cite{Lin93} and the
smoothed Schoenfeld residuals method proposed by
\cite{therneau00}. Both methods agree that the PH assumption seems to
be problematic for the Karnofsky score (Figure
\ref{fig:ovarian2_PH}). The test based on cumulative Schoenfeld residuals  returns
a $p$-value of 0.009. The left hand graph of Figure
\ref{fig:ovarian2_PH} (smoothed residuals) shows a rapid early drop in
importance of the Karnofsky score, implying that baseline Karnofsky
score, measured at diagnosis, is not predictive of mortality beyond
the first year of follow-up, something that could be expected for
advanced cancer. \\

\noindent\emph{Dealing with the lack of PH:}\\
One approach to deal with the violation of proportional hazards
assumption is to fit a set of landmark models. In this case we
consider 2-year windows and set the landmark points at 0, 1 and 2
years. By fitting a separate model at each landmark point, we allow
the coefficient to change in time but also to use the most recent
covariate values for the prediction (in this case there are no
time-dependent covariates, so all fits use the same values). As can be
seen from Table \ref{tab:ovarian2}, the effect of the residual tumor
diameter increases with size at all the landmark points, though quite
some variability can be observed there (the subgroups are rather small
and hence the standard errors quite large). The coefficient for the
stage at baseline (FIGO) remains rather constant in time, whereas the
effect of Karnofsky index quickly approaches zero (hazard ratio approaches 1), for predictions
from 1 or 2 years onward, the baseline value of Karnofsky score
carries no important information. \\

\begin{table} \centering
   \begin{tabular}{l|cc|cc|cc}
& \multicolumn{2}{c}{From time 0} & \multicolumn{2}{|c|}{From 1 year} &\multicolumn{2}{c}{From 2 years}  \\ \hline
&HR& 95\% CI&HR& 95\% CI&HR& 95\% CI\\ \hline
Diameter$<$1cm & 1.31 &(0.5-3.3)&1.63 &(0.7-4.1)&2.73 &(0.8-9.4) \\
Diameter 1-2cm & 2.92 &(1.2-7.1)&2.89 &(1.2-7.2)&2.17 &(0.5-8.7) \\
Diameter 2-5cm & 3.04 &(1.3-7.2)&2.75 &(1.1-6.7)&3.55 &(1.0-12.7) \\
Diameter$>$5cm & 2.69 &(1.2-6.3)&3.21 &(1.4-7.6)&5.54 &(1.7-18.4) \\
FIGO (stage IV vs. III) & 1.76 &(1.3-2.4)&1.70 &(1.2-2.5)&1.64 &(0.9-2.9) \\
Karnofsky index & 0.77 &(0.7-0.9)&0.89 &(0.8-1.0)&1.07 &(0.8-1.4) \\
\end{tabular}
   \caption{Estimated hazard ratios (HR) and 95 \% confidence intervals (CI) in landmark models (with 2-year windows) for the Ovarian cancer data.
   \label{tab:ovarian2}
}
\end{table}

\begin{figure}\centering
  \includegraphics[height=5cm]{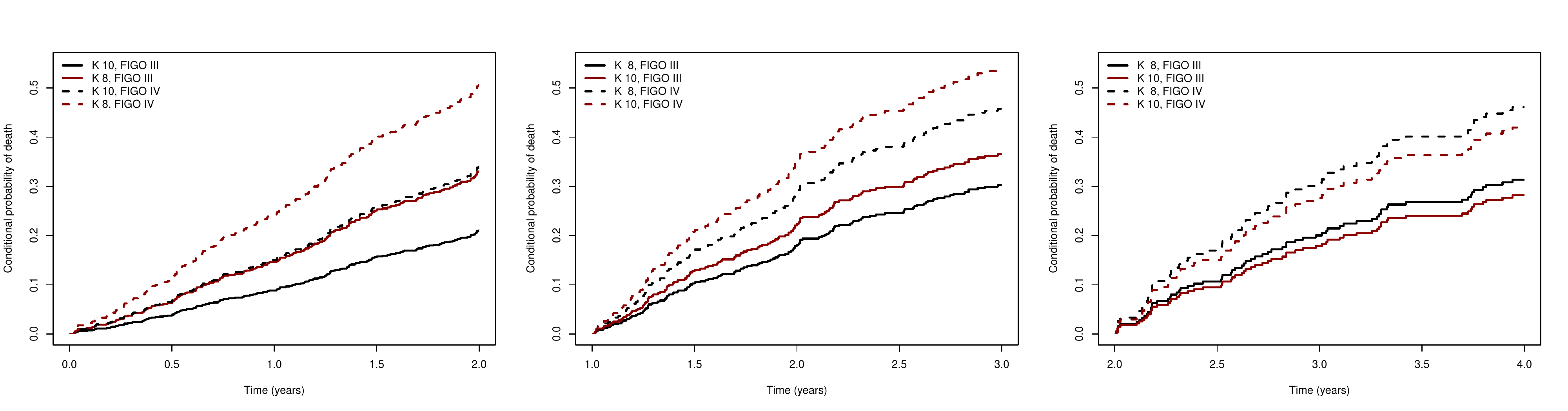}
\caption{The conditional probability of dying for patients with diameter $<1$ cm with respect to stage and two chosen levels of Karnofsky index. Two-year conditional probabilities for patients still at risk at the beginning of each time window are estimated.}
\label{fig:ovarian4_pred}
\end{figure}

\noindent \textit{Prediction:}
The hazard ratios describe the relative effect of each covariate, but
do not tell anything about the absolute risk of the patients. To this
end, Figure \ref{fig:ovarian4_pred} shows risk estimates for a set of
covariate values. The probability of dying in the first 2 years is
comparable in size to the conditional probability of dying in the next
2 years at each of chosen time points. As seen in Table
\ref{tab:ovarian2}, the Karnofsky index at baseline is crucial for the
prognosis in the first 2 years, but less relevant for patients who
survive the initial period. Obviously, observation of an updated Karnofsky index could change this conclusion.

In conclusion, the fitted Cox regression model enabled estimation of
the absolute risk of dying within two years, even in the presence of a
covariate for which the PH assumption was not satisfied. For that
purpose, the technique of landmarking proved very useful.

\section{Summary and discussion}

\label{sec:disc}

In general multi-state models, the \emph{intensity} is the basic
parameter \citep{ABGK-book}
and we have argued that, in the analysis of time-to-event data from
observational studies,
the intensity is, therefore, an obvious parameter to target. Focus has been on a single occurrence
of a single type of event, such as (cause-specific) death, onset/diagnosis of a disease, or first
hospital re-admission. Recurrent themes have been that hazard models known from survival analysis
are applicable in such situations and that studies of this kind have a number of common features.
These include, e.g., specification of the time axis for analysis, how to deal with incomplete observation
in the form of right-censoring and delayed entry, and how to use and interpret models including
time-dependent covariates. Also, the concept of immortal time bias is relevant in all such studies.

We have provided some checklists that we find useful to consider, however, it is important to emphasize
that these checklists cannot be taken as `cook books' on how to conduct time to event analysis in
observational studies. Rather, they are meant as guidelines and we have also emphasized that the most
important item to consider when planning such an analysis is to clearly specify the research question
and think about to what extent the available data allow an answer to that question. We have also identified
research questions for which an intensity model only provides one step towards an answer and where further
analyses are needed. These include risk prediction for non-fatal events and causal inference.

Finally, we have presented some worked examples using the methods summarized in earlier sections and going
through the checklists provided. Our examples illustrate also the need to interpret the results with some caution, taking into account both the limitations of the data at hand and the underlying assumptions, which should be carefully checked, possibly triggering some additional analysis. Further details concerning these examples are collected as Supplementary
Material that also includes information on how the analysis results were obtained using {\tt R}.

Even though the paper is not short, it fails to discuss a number of aspects that are also of importance.
These include most mathematical details about properties of the methods, as well as more comprehensive analyses of data with competing
risks, recurrent events, and more general multi-state models
\citep{cook-lawless-book2007, cook-lawless-book2018}. We have focussed
on the Cox regression model
throughout (and to a lesser extent the piecewise exponential/`Poisson'
model) and discussion of AFT models,
additive hazards models as well as random effects (`frailty') models,
e.g. joint models for the event intensity
and an internal time-dependent covariate, is not included
\citep{hougaard-book, rizo-book}. The same holds for models for
relative survival \citep{maja-etal} and how to deal with
interval-censoring \citep{joly-etal}. Some of these may be topics for
forthcoming papers from the STRATOS TG8 topic group.

\subsection*{Acknowledgements}

MA is a James McGill Professor at McGill University. His research is supported by the Natural Sciences and Engineering Research Council of Canada (NSERC) grant 228203 and the Canadian Institutes of Health Research (CIHR) grant PJT-148946.
The research of MPP is supported by Slovenian Research Agency (grant P3-0154 , 'Methodology for data analysis in medical sciences').

\bibliography{paperref}

\end{document}